\documentclass[%
 aip,
 amsmath,amssymb,
preprint,
]{revtex4-1}

\usepackage{graphicx}
\usepackage{dcolumn}
\usepackage{bm}
\usepackage[utf8]{inputenc}
\usepackage[T1]{fontenc}
\usepackage{mathptmx}
\usepackage{etoolbox}

\usepackage{framed,multirow}
\usepackage[utf8]{inputenc}
\usepackage{bm}
\usepackage{amsmath}
\usepackage[mathscr]{euscript}
\usepackage{graphicx}
\usepackage{epsfig, setspace}
\usepackage{url}
\usepackage{graphicx, transparent, color}
\usepackage{algorithm} 
\usepackage{algorithmicx}
\usepackage{etex}
\usepackage{siunitx}
\usepackage{soul}
 \usepackage[hidelinks]{hyperref}
\usepackage{amsmath}

\usepackage{silence}
\usepackage{ulem,cancel}
\usepackage[utf8]{inputenc}
\newtheorem{remark}{\bf Remark}[section]
\newtheorem{example}{Example}[section]
\usepackage{graphicx}
\usepackage{float}
\usepackage{subfigure}
\usepackage{subfigmat}

\usepackage{setspace,lipsum}
\usepackage{dirtytalk}

\usepackage[top=18truemm,bottom=18truemm,left=20truemm,right=20truemm]{geometry}
\DeclareFontFamily{U}{mathx}{}
\DeclareFontShape{U}{mathx}{m}{n}{<-> mathx10}{}
\DeclareSymbolFont{mathx}{U}{mathx}{m}{n}
\DeclareMathAccent{\widehat}{0}{mathx}{"70}
\DeclareMathAccent{\widecheck}{0}{mathx}{"71}
\usepackage[table]{xcolor}

\makeatletter
\def\@email#1#2{%
 \endgroup
 \patchcmd{\titleblock@produce}
  {\frontmatter@RRAPformat}
  {\frontmatter@RRAPformat{\produce@RRAP{*#1\href{mailto:#2}{#2}}}\frontmatter@RRAPformat}
  {}{}
}%
\makeatother
\begin{document}

\preprint{AIP/123-QED}

\title[]{A Wave Appropriate Discontinuity Sensor Approach for Compressible Flows}

\author{Amareshwara Sainadh Chamarthi}
\thanks{Authors to whom correspondence should be addressed: a.s.chamarthi@gmail.com}

\author{Natan Hoffmann}

\author{Steven Frankel}

\affiliation{Faculty of Mechanical Engineering, Technion - Israel Institute of Technology, Haifa, 3200003, Israel}

\date{\today}

\begin{abstract}

In this work, we propose a novel selective discontinuity sensor approach for numerical simulations of \textcolor{black}{the} compressible Navier-Stokes equations. Since transformation to characteristic space is already a common approach to reduce high-frequency oscillations during interpolation to cell interfaces, we exploit the characteristic wave structure of the Euler equations to selectively treat the various waves that the equations comprise. The approach uses the Ducros shock sensing criterion to detect and limit oscillations due to shocks while applying a different criterion to detect and limit oscillations due to contact discontinuities. Furthermore, the method is general in the sense that it can be applied to any method that employs characteristic transformation and shock sensors. However, in the present work, we focus on the Gradient-Based Reconstruction family of schemes. A series of inviscid and viscous test cases containing various types of discontinuities are carried out. The proposed method is shown to markedly reduce high-frequency oscillations that arise due to improper treatment of the various discontinuities; i.e., applying the Ducros shock sensor in a flow where a strong contact discontinuity is present. Moreover, the proposed method is shown to predict similar volume-averaged kinetic energy and enstrophy profiles for the Taylor-Green vortex simulation compared to the base Ducros sensor, indicating that it does not introduce unnecessary numerical dissipation when there are no contact discontinuities in the flow.

\end{abstract}

\maketitle

\section{\label{sec:introduction}Introduction}

Compressible flows may exhibit various types of discontinuities. For example, whereas shock waves occur when the local flow velocity exceeds the local speed of sound, contact discontinuities can form due to flow streams with different velocities, causing a shear layer, or at material interfaces where two fluids of different densities meet. Capturing these various discontinuities in the context of numerical simulations is challenging since the properties of these discontinuities differ, and their numerical treatment is key to their accurate representation. In addition, this challenge is exacerbated when turbulence is present, as turbulence requires separate numerical treatment altogether. 

The ability to numerically simulate discontinuities and turbulence beckons the use of discontinuity detectors and/or filtering methods. \textcolor{black}{This} is necessary because accurately resolving turbulence requires low numerical dissipation, and this lack of dissipation can cause oscillations and eventual blow-up of the simulation in the presence of discontinuities. Thus, hybrid schemes are often employed for these types of flows, utilizing a low-dissipation numerical scheme in smooth flow regions, sensing regions with discontinuities via some given set of criteria, and applying a discontinuity-capturing scheme in these regions. Consequently, the sensor used to switch between the schemes is very important.

Sensors for shocks have been used rather successfully in the past. One of the first was through the Jameson scheme, wherein dissipation is introduced through the numerical flux for regions where a sensor based on pressure fluctuations detects a shock. This method was coined the Jameson-Schmidt-Turkel artificial flux on account of the authors of the original paper \cite{jameson1981numerical}. Ducros et al. \cite{ducros1999large} extended the Jameson-Schmidt-Turkel artificial flux by multiplying the pressure fluctuation-based sensor by a function of the local dilatation and vorticity. \textcolor{black}{This sensor has been widely employed in the literature \cite{fang2023direct,fang2013optimized,trettel2016mean}}. Moreover, various modifications have been made to improve its capabilities. Pirozzoli \cite{pirozzoli2011numerical} highlighted various shock-detecting methods, including the Ducros shock sensor and further improved it by modifying one of the terms to exclude the sensor in turbulent boundary layer regions. Hendrickson et al. \cite{hendrickson2018improved} improved the Ducros sensor by mitigating its effect in regions of small dilatation. 

The Ducros sensor can be implemented in various fashions. One option is to use a hybrid flux technique where the numerical flux is split into non-dissipative and dissipative sub-components. The dissipative sub-component is pre-multiplied by the Ducros sensor and effectively applies the dissipative flux in regions where the sensor detects a shock. However, this method requires using a skew-symmetric form of the numerical flux and is thus not generally applicable to various numerical methods. Another option is to switch between a low-dissipation and shock-capturing scheme if the Ducros sensor value exceeds some cutoff value. However, this option has the drawback of relying on a case-dependent parameter. Studies have highlighted the effects of varying this cutoff value, such as in De Vanna et al. \cite{deeffect} or van Noordt et al. \cite{van2022immersed}. The reliance of the sensor on a parameter is undoubtedly a disadvantage. However, for now, it is necessary when applying the Ducros sensor to existing methods that do not rely on alternate forms of the numerical flux. 

While the Ducros shock sensor is a well-established method for detecting shocks, other discontinuities can arise in compressible flows that require separate numerical treatment. For example, contact discontinuities arising in compressible flows require their own detection criteria, distinct from those of shock-detection methods, because they form as a result of different physical phenomena to those of shocks. Despite this, a contact discontinuity sensor that effectively treats the immediate regions containing contact discontinuities has received markedly less attention in the literature. 

In this work, we propose a novel selective discontinuity sensor approach based on the characteristic waves of the Euler equations. This approach exploits the Euler wave structure, which comprises acoustic waves, an entropy wave, and shear/vorticity waves. By employing discontinuity sensors depending on the characteristic wave type, one can overcome the inability of the Ducros sensor to detect discontinuities other than shock waves.

\textcolor{black}{There are a vast amount of numerical methods designed for capturing discontinuities \cite{li2021fifth,yu2023weighted}}. Since no sensor is perfect, the choice of discontinuity-capturing method is also very important since excessive dissipation in regions not containing discontinuities can severely affect the solution. On top of that, discontinuity-capturing methods are generally more computationally intensive than smooth flow schemes. Therefore, the choice of this scheme must be made with care. Of the vast amount of discontinuity-detecting schemes, the weighted-essentially-non-oscillatory (WENO) scheme introduced by Liu et al. \cite{liu1994weighted} is the most widely used. This scheme has received much attention since its inception in 1994, with many modifications proposed to improve it's efficacy. Some examples include the improvements made by Jiang and Shu \cite{jiang1996efficient} and, more recently, the targeted-ENO (TENO) family of schemes introduced by Fu et al. \cite{fu2016family}. While WENO and TENO schemes have proven to be robust methods for shock-capturing capable of application to various types of flows, they can be over-dissipative and computationally expensive. Monotonicity-preserving (MP) methods such as the MP5 scheme of Suresh and Hyunh \cite{suresh1997accurate} have also been used for flows with discontinuities. MP schemes employ a limiter to limit high-frequency oscillations caused by discontinuities. While they have proven to be robust and less computationally demanding than the WENO family of schemes, improvements can still be made, as MP limiters can become too sensitive and cause excessive dissipation. It should also be noted that WENO schemes, specifically very high-order schemes, often use the MP approach to prevent oscillations \cite{balsara2000monotonicity,li2021low} and improve robustness. In a recent paper, Li et al. \cite{li2023family} mentioned that the smoothness indicators used in the ENO-type schemes for shock-capturing are very expensive and proposed a regularization approach based on the MP criterion. Another hybrid approach recently gaining more attention is the Boundary Variation Diminishing (BVD) algorithm of Sun et al. \cite{sun2016boundary}. The BVD algorithm combines a smooth flow and shock-capturing scheme, switching between the two by comparing each scheme's total boundary variation (TBV) at cell interfaces. In this fashion, the BVD algorithm chooses the least oscillatory interpolation and minimizes the amount of numerical dissipation at cell interfaces. This principle may combine virtually any smooth flow and shock-capturing scheme. Recently, Chamarthi and Frankel \cite{chamarthi2021high} introduced an adaptive central-upwind scheme, which employs a sixth-order linear-compact scheme and the fifth-order MP scheme, using the BVD algorithm to switch between the two candidates. While the method shows promise, the BVD algorithm may also lead to excessive dissipation for turbulent flow simulations and is expensive as two different schemes must be used \cite{chamarthi2023implicit}.

More recently, the Gradient-Based Reconstruction (GBR) approach, introduced by Chamarthi \cite{chamarthimig2022}, was developed to address the niche of stable, high-accuracy numerical methods with excellent spectral properties for smooth and discontinuous flows. In general, this family of schemes employs the first two moments of the Legendre basis to interpolate the numerical fluxes from the cell center to the cell interface. Since both the first and second derivatives of a given flow variable are required for this interpolation, the scheme has the added advantage of computing high-order gradients that can be re-used throughout the solver and for post-processing. In other words, the computed gradients may be used for the inviscid fluxes, viscous fluxes, discontinuity detector, and post-process quantities such as enstrophy or Q-criterion. The distinction from one GBR method to another can generally be attributed to the method that is used to evaluate the first and second derivatives in the interpolation. For instance, if compact finite differences are used for these derivatives, the scheme is designated an Implicit Gradient (IG) method. In contrast, if explicit finite differences are used, the scheme is aptly named an Explicit Gradient (EG) method. This family of schemes was extended to handle discontinuous flows by applying the MP limiter of Suresh and Hyunh \cite{suresh1997accurate} to the numerical flux interpolation, which was found to be both robust and exhibit good spectral properties. The MP-limited GBR methods are designated MIG and MEG. While MIG and MEG were found to perform well against other widely used schemes, in the context of turbulent flows, the MP limiter may cause excessive dissipation. Thus, an improvement must be addressed.

As such, this work has two objectives. The first is to extend the GBR method to handle compressible turbulent flows by employing the Ducros shock sensor to mitigate the cited excessive dissipation. The second objective is to overcome the cited inability of the Ducros sensor to detect contact discontinuities by proposing a selective discontinuity detector based on characteristic waves that can appropriately detect and capture both shock waves and contact discontinuities.

The rest of the paper is organized as follows. In the next section, the governing equations are presented. After that, the numerical method is explained in detail, and the proposed characteristic wave appropriate discontinuity detector is delineated. Following that, results from a set of test cases are shown and discussed. Lastly, concluding remarks are made, and future work is set forth.

\section{\label{sec:governingEquations}Governing Equations}

In this study, the three-dimensional compressible Navier-Stokes equations in conservative form are solved in Cartesian coordinates:

\begin{equation}\label{eqn:cns}
    \centering
    \frac{\partial \mathbf{U}}{\partial t} + \frac{\partial \mathbf{F}^c}{\partial x} + \frac{\partial \mathbf{G}^c}{\partial y} + \frac{\partial \mathbf{H}^c}{\partial z} + \frac{\partial \mathbf{F}^v}{\partial x} + \frac{\partial \mathbf{G}^v}{\partial y} + \frac{\partial \mathbf{H}^v}{\partial z} = 0,
\end{equation}

\noindent where $t$ is time and $(x,y,z)$ are the Cartesian coordinates. $\mathbf{U}$ is the conserved variable vector, and $\mathbf{F}^c$, $\mathbf{G}^c$, and $\mathbf{H}^c$ are the convective flux vectors defined as:

\begin{subequations}
    \centering
    \begin{gather}
        \mathbf{U} = \begin{pmatrix}
        \rho \\
        \rho u \\
        \rho v \\
        \rho w \\
        \rho E
        \end{pmatrix},
        \quad
        \mathbf{F}^c = \begin{pmatrix}
        \rho u \\
        \rho u u + p \\
        \rho u v + p \\
        \rho u w + p \\
        \rho u H 
        \end{pmatrix},
        \quad
        \mathbf{G}^c = \begin{pmatrix}
        \rho v \\
        \rho v u + p \\
        \rho v v + p \\
        \rho v w + p \\
        \rho v H 
        \end{pmatrix},
        \quad
        \mathbf{H}^c = \begin{pmatrix}
        \rho w \\
        \rho w u + p \\
        \rho w v + p \\
        \rho w w + p \\
        \rho w H 
        \end{pmatrix},
        \tag{\theequation a--\theequation d}
    \end{gather}
\end{subequations}

\noindent where $\rho$ is density, $u$, $v$, and $w$ are the velocities in the $x$, $y$, and $z$ directions, respectively, $p$ is the pressure, $E = e + \left(u^2 + v^2 + w^2 \right)/2$ is the specific total energy, and $H = E + p/\rho$ is the specific total enthalpy. The equation of state is for a calorically perfect gas so that $e = p/ \left[ \rho (\gamma-1) \right]^{-1}$ is the internal energy, where $\gamma = \mathrm{c_p}/\mathrm{c_v}$ is the ratio of specific heats, with $\mathrm{c_p}$ as the isobaric specific heat and $\mathrm{c_v}$ as the isochoric specific heat. $\mathbf{F}^v$, $\mathbf{G}^v$, and $\mathbf{H}^v$ are the viscous flux vectors defined as:

\begin{subequations}
    \centering
    \begin{gather}
        \mathbf{F}^v = - \begin{pmatrix}
        0 \\
        \tau_{xx} \\
        \tau_{xy} \\
        \tau_{xz} \\
        \tau_{xx} u + \tau_{xy} v + \tau_{xz} w - q_x
        \end{pmatrix},
        \tag{\theequation a--\theequation b}
        \quad
        \mathbf{G}^v = - \begin{pmatrix}
        0 \\
        \tau_{yx} \\
        \tau_{yy} \\
        \tau_{yz} \\
        \tau_{yx} u + \tau_{yy} v + \tau_{yz} w - q_y
        \end{pmatrix},
        \\[15pt]
        \mathbf{H}^v = - \begin{pmatrix}
        0 \\
        \tau_{zx} \\
        \tau_{zy} \\
        \tau_{zz} \\
        \tau_{zx} u + \tau_{zy} v + \tau_{zz} w - q_z
        \end{pmatrix},
        \tag{\theequation c}
    \end{gather}
\end{subequations}

\noindent where the normal stresses are:

\begin{subequations}
    \centering
    \begin{gather}
        \tau_{xx} = 2 \hat{\mu} \frac{\partial u}{\partial x} + \hat{\lambda} \left(\frac{\partial u}{\partial x} + \frac{\partial v}{\partial y} + \frac{\partial w}{\partial z} \right),
        \tag{\theequation a--\theequation b}
        \quad
        \tau_{yy} = 2 \hat{\mu} \frac{\partial v}{\partial y} + \hat{\lambda} \left(\frac{\partial u}{\partial x} + \frac{\partial v}{\partial y} + \frac{\partial w}{\partial z} \right), 
        \\[15pt]
        \tau_{zz} = 2 \hat{\mu} \frac{\partial w}{\partial z} + \hat{\lambda} \left(\frac{\partial u}{\partial x} + \frac{\partial v}{\partial y} + \frac{\partial w}{\partial z} \right),
        \tag{\theequation c}
    \end{gather}
\end{subequations}

\noindent where $\hat{\mu} = \mu/\mathrm{Re}$ is the scaled dynamic viscosity as a result of non-dimensionalization and Stokes' hypothesis is assumed so that $\hat{\lambda} = -\frac{2}{3} \hat{\mu}$. The dynamic viscosity is a function of temperature by Sutherland's law:

\begin{equation}
    \mu(T) = T^{3/2} \frac{1 + S/T_{ref}}{T + S/T_{ref}},
    \label{eqn:sutherlandsLaw}
\end{equation}

\noindent where $T_{ref}$ is the reference temperature and $S = \SI{110.4}{\kelvin}$ is Sutherland's constant. $\mathrm{Re} = \rho_{\infty} u_{\infty} L_{ref}/\mu_{\infty}$ is the Reynolds number where the $( \cdot )_{\infty}$ subscript denotes a freestream value. The shear stresses are defined as:

\begin{subequations}
    \centering
    \begin{gather}
        \tau_{xy} = \tau_{yx} = \hat{\mu} \left(\frac{\partial u}{\partial y} + \frac{\partial v}{\partial x} \right),
        \quad
        \tau_{yz} = \tau_{zy} = \hat{\mu} \left(\frac{\partial v}{\partial z} + \frac{\partial w}{\partial y} \right),
        \quad
        \tau_{xz} = \tau_{zx} = \hat{\mu} \left(\frac{\partial u}{\partial z} + \frac{\partial w}{\partial x} \right),
        \tag{\theequation a--\theequation c}
    \end{gather}
\end{subequations}

\noindent and the heat fluxes are:

\begin{subequations}
    \centering
    \begin{gather}
        q_{x} = -\hat{\kappa} \frac{\partial T}{\partial x},
        \quad
        q_{y} = -\hat{\kappa} \frac{\partial T}{\partial y},
        \quad
        q_{z} = -\hat{\kappa} \frac{\partial T}{\partial z},
        \tag{\theequation a--\theequation c}
    \end{gather}
\end{subequations}

\noindent where $\hat{\kappa} = \mu \left( \mathrm{Ma}^2 \mathrm{Re}(\gamma-1)\mathrm{Pr} \right)^{-1}$ is the scaled thermal conductivity, $\mathrm{Ma} = u_{\infty} \left( \gamma R_{gas} T \right)^{-1/2}$ is the Mach number, $\mathrm{Pr}$ is the Prandtl number, $T$ is the temperature, and $R_{gas}$ is the universal gas constant. The equations are non-dimensionalized using the freestream density $\rho_{\infty}$, the freestream velocity $u_{\infty}$, reference length $L_{ref}$, the freestream temperature $T_{\infty}$, and the freestream dynamic viscosity $\mu_{\infty}$ such that the temperature is related to pressure and density via $ p = \rho T \left( \gamma \mathrm{Ma}^2 \right)^{-1}$.

\section{\label{sec:numericalMethods}Numerical Methods}

In this work, we extend the GBR method of Chamarthi \cite{chamarthimig2022} by employing the Ducros shock sensor to more sensitively detect and limit oscillations arising from shocks. \textcolor{black}{Furthermore}, we propose a selective discontinuity detector to effectively capture both shock waves and contact discontinuities. In what follows, we will delineate the details of this method. After, we will briefly present the viscous flux discretization and the time integration method.

\subsection{Convective Flux Spatial Discretization Scheme}

Using a conservative numerical method, the governing equations cast in semi-discrete form for a Cartesian cell $I_{i,j,k} = \left[ x_{i-\frac{1}{2}}, x_{i+\frac{1}{2}} \right] \times \left[ y_{i-\frac{1}{2}}, y_{i+\frac{1}{2}} \right] \times \left[ z_{i-\frac{1}{2}}, z_{i+\frac{1}{2}} \right]$ can be expressed via the following ordinary differential equation: 

\begin{align}
    \begin{aligned}
        \frac{\text{d}}{\text{d} t} \check{\mathbf{U}}_{i,j,k} = \mathbf{Res}_{i,j,k} = &- \left. \frac{\text{d} \check{\mathbf{F}}^c}{\text{d} x} \right|_{i,j,k} - \left. \frac{\text{d} \check{\mathbf{G}}^c}{\text{d} y} \right|_{i,j,k} - \left. \frac{\text{d} \check{\mathbf{H}}^c}{\text{d} z} \right|_{i,j,k} \\ 
        &+ \left. \frac{\text{d} \check{\mathbf{F}}^v}{\text{d} x} \right|_{i,j,k} + \left. \frac{\text{d} \check{\mathbf{G}}^v}{\text{d} y} \right|_{i,j,k} + \left. \frac{\text{d} \check{\mathbf{H}}^v}{\text{d} z} \right|_{i,j,k},
    \end{aligned}
\end{align}

\noindent where the check accent, $\check{(\cdot)}$, indicates a numerical approximation of a physical quantity, $\mathbf{Res}_{i,j,k}$ is the residual function, and the remaining terms are cell center numerical flux derivatives of the physical fluxes in Equation (\ref{eqn:cns}). For brevity, we continue with only the $x$-direction, however, the following may be extended to all three dimensions straightforwardly. Moreover, we drop the $j$ and $k$ indices in the interest of clarity. The cell center numerical convective flux derivative is expressed as:

\begin{equation}
    \left. \frac{\text{d} \check{\mathbf{F}}^c}{\text{d} x} \right|_{i} = \frac{1}{\Delta x} \left( \check{\mathbf{F}}^{c}_{i+\frac{1}{2}} - \check{\mathbf{F}}^{c}_{i-\frac{1}{2}} \right),
\end{equation}

\noindent where $i \pm \frac{1}{2}$ indicates right and left cell interface values, respectively. Using a GBR method, $\check{\mathbf{F}}^c_{i \pm \frac{1}{2}}$ are computed using an approximate Riemann solver, since a Riemann problem exists at each cell interface. The interface numerical convective fluxes are computed from:

\begin{equation}
    \check{\mathbf{F}}^c_{i \pm \frac{1}{2}} = \frac{1}{2} \left[ \check{\mathbf{F}}^c \left( \check{\mathbf{U}}^{L}_{i \pm \frac{1}{2}} \right) + \check{\mathbf{F}}^c \left( \check{\mathbf{U}}^{R}_{i \pm \frac{1}{2}} \right) \right] - \frac{1}{2} \left| \mathbf{A}_{i \pm \frac{1}{2}} \right| \left( \check{\mathbf{U}}^{R}_{i \pm \frac{1}{2}} - \check{\mathbf{U}}^{L}_{i \pm \frac{1}{2}} \right),
\end{equation}

\noindent where the $L$ and $R$ superscripts denote the left- and right-biased states, respectively, and $\left| \mathbf{A}_{i \pm \frac{1}{2}} \right|$ denotes the convective flux Jacobian. In this work, the Hartex-Lax-van Leer-Contact (HLLC) \cite{toro2009riemann} approximate Riemann solver is used unless otherwise explicitly stated. The objective is to obtain the left- and right-biased states. These are computed with the GBR method, which will be explained in the following subsection.

\subsubsection{Gradient-Based Reconstruction Method: Linear Scheme}

GBR methods employ the first two moments of the Legendre polynomial evaluated on $x_{i-\frac{1}{2}} \leq x \leq x_{i+\frac{1}{2}}$ for interpolation. This may be written as:

\begin{equation}\label{eqn:legendre}
    \mathbf{U}(x) = \mathbf{U}_{i} + \frac{\mathbf{U}'_{i}}{\Delta x} (x-x_i) + \frac{3 \mathbf{U}''_{i}}{2 \Delta x^{2}_{i}} \mathscr{K} \left[ (x-x_i)^{2} - \frac{\Delta x^{2}_{i}}{12} \right], 
\end{equation}

\noindent where $\mathbf{U}'_{i}$ and $\mathbf{U}''_{i}$ respectively represent the first and second derivatives of the conservative variables $\mathbf{U}$. If $x = x_i + \Delta x/2$ and $\mathscr{K} = 1/3$, the following equations for the left- and right-biased states are obtained:
    
\begin{subequations}
    \begin{gather}
        \mathbf{U}^{L}_{i+\frac{1}{2}} = \mathbf{U}_{i} + \frac{1}{2} \mathbf{U}'_{i} + \frac{1}{12} \mathbf{U}''_{i},
        \quad
        \mathbf{U}^{R}_{i+\frac{1}{2}} = \mathbf{U}_{i+1} - \frac{1}{2}\mathbf{U}i'_{i+1} + \frac{1}{12} \mathbf{U}''_{i+1}.
        \tag{\theequation a--\theequation b}
    \end{gather}
    \label{eqn:legendreInterpolation}
\end{subequations}

\noindent Since in this work the IG scheme is used, $\mathbf{U}'_{i}$ is computed using compact finite differences:
    
\begin{equation}
    \frac{5}{14} \mathbf{U}'_{i-1} + \mathbf{U}'_{i} + \frac{5}{14} \mathbf{U}'_{i+1} = \frac{1}{28 \Delta x} \left( \mathbf{U}_{i+2} - \mathbf{U}_{i-2} \right) + \frac{11}{14 \Delta x} \left( \mathbf{U}_{i+1} - \mathbf{U}_{i-1} \right),
    \label{eqn:firstDerivative}
\end{equation}

\noindent where the coefficients correspond to an optimized fourth-order scheme. $\mathbf{U}''_{i}$ is computed from \cite{chamarthimig2022}:

\begin{equation}
    \mathbf{U}''_{i} = \frac{2}{\Delta x^2} \left( \mathbf{U}_{i+1} - 2 \mathbf{U}_{i} + \mathbf{U}_{i-1} \right) - \frac{1}{2 \Delta x} \left( \mathbf{U}'_{i+1} - \mathbf{U}'_{i-1} \right).
    \label{eqn:secondDerivative}
\end{equation}

\subsubsection{Gradient-Based Reconstruction Method: Non-Linear Scheme}

Eqns. \ref{eqn:legendreInterpolation} are linear interpolations. Therefore, they may be susceptible to oscillations in the presence of discontinuities. So, MP limiting is employed as in Chamarthi \cite{chamarthimig2022}. The following delineates the MP limiting procedure for the left-biased state, however, the procedure is the same for the right-biased state. The MP limiting criterion is:

\begin{equation}
    \mathbf{U}^{L}_{i+\frac{1}{2}} = 
    \begin{cases}
        \mathbf{U}^{L,Linear}_{i+\frac{1}{2}} & \text{if } \left( \mathbf{U}^{L,Linear}_{i+\frac{1}{2}} - \mathbf{U}_i \right) \left( \mathbf{U}^{L,Linear}_{i+\frac{1}{2}} - \mathbf{U}^{L,MP}_{i+\frac{1}{2}} \right) \leq 10^{-20}, \\[5pt]
        \mathbf{U}^{L,Non-Linear}_{i+\frac{1}{2}} & \text{otherwise},
    \end{cases}
    \label{eqn:mpLimitingCriterion}
\end{equation}

\noindent where $\mathbf{U}^{L,Linear}_{i+\frac{1}{2}}$ corresponds to Equation (\ref{eqn:legendreInterpolation}a), and the remaining terms are:

\begin{subequations}
    \begin{alignat}{2}
        &\mathbf{U}^{L,Non-Linear}_{i+\frac{1}{2}} &&= \mathbf{U}^{L,Linear}_{i+\frac{1}{2}} + \text{minmod} \left( \mathbf{U}^{L,MIN}_{i+\frac{1}{2}} - \mathbf{U}^{L,Linear}_{i+\frac{1}{2}}, \mathbf{U}^{L,MAX}_{i+\frac{1}{2}} - \mathbf{U}^{L,Linear}_{i+\frac{1}{2}} \right),
        \\[5pt]
        &\mathbf{U}^{L,MP}_{i+\frac{1}{2}} &&= \mathbf{U}^{L,Linear}_{i+\frac{1}{2}} + \text{minmod} \left[ \mathbf{U}_{i+1}-\mathbf{U}_{i}, \mathscr{A} \left( \mathbf{U}_{i}-\mathbf{U}_{i-1} \right) \right], 
        \\[5pt]
        &\mathbf{U}^{L,MIN}_{i+\frac{1}{2}} &&= \max \left[ \min \left( \mathbf{U}_{i}, \mathbf{U}_{i+1}, \mathbf{U}^{L,MD}_{i+\frac{1}{2}} \right), \min \left( \mathbf{U}_{i}, \mathbf{U}^{L,UL}_{i+\frac{1}{2}}, \mathbf{U}^{L,LC}_{i+\frac{1}{2}} \right) \right],
        \\[5pt]
        &\mathbf{U}^{L,MAX}_{i+\frac{1}{2}} &&= \min \left[ \max \left( \mathbf{U}_{i}, \mathbf{U}_{i+1}, \mathbf{U}^{L,MD}_{i+\frac{1}{2}} \right), \max \left( \mathbf{U}_{i}, \mathbf{U}^{L,UL}_{i+\frac{1}{2}}, \mathbf{U}^{L,LC}_{i+\frac{1}{2}} \right) \right],
        \\[5pt]
        &\mathbf{U}^{L,MD}_{i+\frac{1}{2}} &&= \frac{1}{2} \left( \mathbf{U}_{i} + \mathbf{U}_{i+1} \right) - \frac{1}{2} d^{L,M}_{i+\frac{1}{2}},
        \\[5pt]
        &\mathbf{U}^{L,UL}_{i+\frac{1}{2}} &&= \mathbf{U}_{i} + 4 \left( \mathbf{U}_{i} - \mathbf{U}_{i-1} \right),
        \\[5pt]
        &\mathbf{U}^{L,LC}_{i+\frac{1}{2}} &&= \frac{1}{2} \left( 3 \mathbf{U}_{i} - \mathbf{U}_{i-1} \right) + \frac{4}{3} d^{L,M}_{i-\frac{1}{2}},
        \\[5pt]
        &d^{L,M}_{i+\frac{1}{2}} &&= \text{minmod} \left( d_i, d_{i+1} \right),
        \\[5pt]
        &d_i &&= 2 \left( \mathbf{U}_{i+1} - 2\mathbf{U}_{i} + \mathbf{U}_{i-1} \right) - \frac{\Delta x}{2} \left( \mathbf{U}'_{i+1} - \mathbf{U}'_{i-1} \right),
        \label{mp_improve}
    \end{alignat}
\end{subequations}

\noindent where $\mathscr{A} = 4$ and $\text{minmod} \left( a,b \right) = \frac{1}{2} \left[ \text{sgn}(a) + \text{sgn}(b) \right] \min \left( \left| a \right|, \left| b \right| \right)$. For the remainder of this work, the non-linear scheme based on the MP limiter will be referred to as MIG (Monotonicity-preserving Implicit Gradient \cite{chamarthimig2022}) and the linear scheme will be denoted as IG4H, similar to that of \cite{chamarthi2023implicit,chamarthimig2022}.

\subsubsection{Ducros Shock Sensor}

While MP limiting effectively mitigates oscillations arising from discontinuities, the detection algorithm in Equation (\ref{eqn:mpLimitingCriterion}) can become too sensitive and cause excessive dissipation. To remedy this issue, the Ducros shock sensor, which is designed specifically to sense shocks, can be used \cite{fang2013optimized}:

\begin{equation}
    \Omega_{i} = \theta_i \frac{ \left( \nabla \cdot \mathbf{u} \right)^2}{ \left( \nabla \cdot \mathbf{u} \right)^2 + \left| \nabla \times \mathbf{u} \right|^2},
    \label{eqn:ducros}
\end{equation}

\noindent where,


\begin{equation}
    \theta_i = \frac{\left| -p_{i-2} + 16 p_{i+1} - 30 p_{i} + 16 p_{i+1} - p_{i+2} \right|}{\left| p_{i-2} + 16 p_{i+1} + 30 p_{i} + 16 p_{i+1} + p_{i+2} \right|},
    \label{eqn:jamesonSensor}
\end{equation}

\noindent and $\mathbf{u}$ is the velocity vector.  We modify $\Omega_{i}$ by using it's maximum value in a three cell neighborhood:

\begin{equation}
    \Omega_{i} = \max \left( \Omega_{i+m} \right), \quad \text{for } m = -1,0,1.    
\end{equation}

\noindent Using $\Omega_i$, Equation (\ref{eqn:mpLimitingCriterion}) is modified to:

\begin{equation}
    \mathbf{U}^{L}_{i+\frac{1}{2}} = 
    \begin{cases}
        \mathbf{U}^{L,Linear}_{i+\frac{1}{2}} & \text{if } \Omega_i \leq 0.01, \\[5pt]
        \mathbf{U}^{L,Non-Linear}_{i+\frac{1}{2}} & \text{otherwise}.
    \end{cases}
    \label{eqn:ducrosLimitingCriterion}
\end{equation}

\noindent For the remainder of this work, the nonlinear shock-capturing scheme that uses the standard Ducros sensor for detecting the discontinuities is denoted MIG-D (D for Ducros). With this method, shocks are detected well and the non-linear scheme effectively limits oscillations. However, since Equation (\ref{eqn:ducrosLimitingCriterion}) only applies the non-linear scheme in regions near shocks, oscillations arising from contact discontinuities can \textit{still} arise.

\subsubsection{Wave Appropriate Discontinuity Sensor}

In this work, we propose a novel selective discontinuity sensor approach involving characteristic variables of the Euler equations. For coupled hyperbolic equations like the Euler equations, shock-capturing should be carried out using characteristic variables for \textit{cleanest} results \cite{van2006upwind}. The algorithm takes advantage of the transformation from physical to characteristic space. The novel algorithm is explained below:

\begin{enumerate}
    
    \item Compute Roe-averaged variables following Blazek \cite{blazek2015computational} (Equation 4.89) to construct the left, $\mathbf{L}_n$, and right, $\mathbf{R}_n$, eigenvectors of the normal convective flux Jacobian. \\
    
    \item Since in this work we use the conservative variables, transform $\check{\mathbf{U}}_{i}$, $\check{\mathbf{U}}'_{i}$, and $\check{\mathbf{U}}''_{i}$ to characteristic space by multiplying them by $\mathbf{L}_n$:

    \begin{subequations}
        \begin{gather}
            \check{\mathbf{C}}_{i+m,b} = \mathbf{L}_{n,i+\frac{1}{2}} \check{\mathbf{U}}_{i+m}, 
            \tag{\theequation a--\theequation b}
            \quad
            \check{\mathbf{C}}'_{i+m,b} = \mathbf{L}_{n,i+\frac{1}{2}} \check{\mathbf{U}}'_{i+m},
            \\[10pt]
            \check{\mathbf{C}}''_{i+m,b} = \mathbf{L}_{n,i+\frac{1}{2}} \check{\mathbf{U}}''_{i+m}, 
            \tag{\theequation c}
        \end{gather}
    \end{subequations}

    for $m = -2,-1,0,1,2,3$ and $b = 1,2,3,4,5$, representing the vector of characteristic variables. For $\check{\mathbf{C}}_{i+m,b}$, in matrix-vector form, this is equivalent to:

    \begin{equation} \label{eqn:blahblah}
        \left( \begin{array}{c}
            \check{\mathbf{C}}_{i+m,1} \\[5pt]
            \check{\mathbf{C}}_{i+m,2} \\[5pt]
            \check{\mathbf{C}}_{i+m,3} \\[5pt]
            \check{\mathbf{C}}_{i+m,4} \\[5pt]
            \check{\mathbf{C}}_{i+m,5}
        \end{array} \right) 
        = 
        \left( \begin{array}{ccccc}
            \dfrac{K q^2}{4 c^2} + \dfrac{q_n}{2 c} & -\left( \dfrac{K}{2 c^2}u + \dfrac{n_x}{2 c} \right) & -\left( \dfrac{K}{2 c^2}v + \dfrac{n_y}{2 c} \right) & -\left( \dfrac{K}{2 c^2}w + \dfrac{n_z}{2 c} \right) & \dfrac{K}{2 c^2} \\[5pt]
            1-\dfrac{K q^2}{2 c^2} & \dfrac{K u}{c^2} & \dfrac{K v}{c^2} & \dfrac{K w}{c^2} & -\dfrac{K}{c^2} \\[5pt]
            -q_l & l_x & l_y & l_z & 0 \\[5pt]
            -q_m & m_x & m_y & m_z & 0 \\[5pt]
            \dfrac{K q^2}{4 c^2} - \dfrac{q_n}{2 c} & -\left( \dfrac{K}{2 c^2}u - \dfrac{n_x}{2 c} \right) & -\left( \dfrac{K}{2 c^2}v - \dfrac{n_y}{2 c} \right) & -\left( \dfrac{K}{2 c^2}w - \dfrac{n_z}{2 c} \right) & \dfrac{K}{2 c^2}
        \end{array} \right)
        \left( \begin{array}{c}
            \check{\mathbf{U}}_{i+m,1} \\[5pt]
            \check{\mathbf{U}}_{i+m,2} \\[5pt]
            \check{\mathbf{U}}_{i+m,3} \\[5pt]
            \check{\mathbf{U}}_{i+m,4} \\[5pt]
            \check{\mathbf{U}}_{i+m,5}
        \end{array} \right)
    \end{equation}

    where $K = \gamma-1$, $c = \sqrt{\gamma p/\rho}$ is the local speed of sound, $q^2 = u^2 + v^2 + w^2$, and $q_n = u n_x + v n_y + w n_z$. In the $x$-direction, $n_x = 1$, whereas $n_y=n_z=0$. The $y$- and $z$-directions are analogous. $\mathbf{l} = \left[ l_x,l_y,l_z \right]^{\text{T}}$, $\mathbf{m} = \left[ m_x,m_y,m_z \right]^{\text{T}}$, and $\mathbf{n}= \left[ n_x,n_y,n_z \right]^{\text{T}}$ are mutually orthogonal unit vectors where ${\text{T}}$ denotes transpose; and $q_l = u l_x + v l_y + w l_z$, $q_m = u m_x + v m_y + w m_z$. For more details, see Masatsuka \cite{masatsuka2013like}. \\

    The second characteristic variable, $\check{\mathbf{C}}_{i+m,2}$, corresponds to what is known in one-dimension as the entropy wave. It is this wave that requires limiting in the presence of contact discontinuities, which significantly improves solution quality in a manner corresponding to the actual physical characteristic of them. \\

    \item Using Equations (\ref{eqn:legendreInterpolation}), obtain the unlimited interpolation to cell interfaces in characteristic space via:

\begin{subequations}
    \begin{gather}
        \check{\mathbf{C}}^{L}_{i+\frac{1}{2},b} = \check{\mathbf{C}}_{i,b} + \frac{1}{2} \check{\mathbf{C}}'_{i,b} + \frac{1}{12} \check{\mathbf{C}}''_{i,b},
        \quad
        \check{\mathbf{C}}^{R}_{i+\frac{1}{2},b} = \check{\mathbf{C}}_{i+1,b} - \frac{1}{2} \check{\mathbf{C}}'_{i+1,b} + \frac{1}{12} \check{\mathbf{C}}''_{i+1,b}.
        \tag{\theequation a--\theequation b}
    \end{gather}
    \label{eqn:unlimitedCharacteristicInterpolation}
\end{subequations}

\noindent The left biased interpolation is then treated by the following algorithm:

\begin{equation}
    \check{\mathbf{C}}^{L}_{i+\frac{1}{2},b} = 
    \begin{cases}
        \check{\mathbf{C}}^{L,Non-Linear}_{i+\frac{1}{2},b} & \text{if } b = 2 \text{ and } 2 \xi/(1 + \xi)^2 \geq 0.01, \quad \text{where } \xi = \left| {\check{\mathbf{C}}}_{i,2} \right|/ \left( \rho_i + \rho_{i+1} \right), 
        \\[10pt]
        \check{\mathbf{C}}^{L,Non-Linear}_{i+\frac{1}{2},b} & \text{if } b \neq 2 \text{ and } \Omega_i > 0.01,
        \\[10pt]
        \check{\mathbf{C}}^{L,Linear}_{i+\frac{1}{2},b} & \text{otherwise}.
    \end{cases}
    \label{eqn:newSensorCriterion}
\end{equation}

The criterion used for $b = 2$ is that of Frahan et al. \cite{de2015new}. It can detect contact discontinuities as it depends on the density. The important idea here is that the contact discontinuity sensor of Frahan et al. \cite{de2015new} can detect contact discontinuities but is over-dissipative if used as a sensor for other flow features. On the other hand, the Ducros sensor can detect shocks and performs well for turbulent flow simulations but cannot detect contact discontinuities. However, one can reasonably overcome each sensor's deficiencies by selectively choosing the sensors according to the Euler equations' wave structure.

 \begin{remark}
    It is also possible to use the MP criterion given by the Equation (\ref{eqn:mpLimitingCriterion}) to detect the contact discontinuities instead of Frahan's detector as in \cite{hoffmann2023large}. It has been observed that the MP criterion is slightly more dissipative than that of Frahan. The proposed selective detector is not limited to the MIG scheme (and MP-type limiting approach) but can also be used along with the ENO-type schemes. Results using the WENO-Z scheme \cite{Borges2008,fu2019low} in lieu of the present non-linear scheme are shown in Appendix A of this manuscript. 
 \end{remark}
 
 \begin{remark}
  The proposed approach is also different from that of Johnsen et al. \cite{johnsen2013recovery}. Johnsen et al. \cite{johnsen2013recovery} flagged the regions using a separate sensor for the contact discontinuities, shock waves, and turbulence and then applied their limiter approach. In the current work, shocks and turbulence are detected by the Ducros sensor itself. \end{remark}

\item After obtaining $\check{\mathbf{C}}^{L,R}_{i+\frac{1}{2},b}$, the reconstructed states are then recovered by projecting the characteristic variables back to physical fields:

\begin{equation}
    \check{\mathbf{U}}^{L,R}_{i+\frac{1}{2}} = \mathbf{R}_{n,i+\frac{1}{2}} \check{\mathbf{C}}^{L,R}_{i+\frac{1}{2}}.
\end{equation}

For the remainder of this work, the nonlinear shock-capturing scheme that uses the selective sensor (Equation (\ref{eqn:newSensorCriterion})) is denoted MIG-S (S for selective). A summary of the presented methods along with the discontinuity detection criterion are shown in Table \ref{tab:all_schemes}.

\begin{table}[H]
    \centering
    \caption{Discontinuity detection criterion of various schemes considered in this paper.}
    \begin{tabular}{ c c c c }
        \hline
        \hline
        Scheme & Criterion \\
        \hline
        MIG   & Equation (\ref{eqn:mpLimitingCriterion}) \\
        MIG-D & Equation (\ref{eqn:ducrosLimitingCriterion}) \\
        MIG-S & Equation (\ref{eqn:newSensorCriterion}) \\
        \hline
        \hline 
    \end{tabular}
    \label{tab:all_schemes}
\end{table}

\end{enumerate}

\subsection{Viscous Flux Spatial Discretization Scheme}

In this subsection, we describe the spatial discretization of the numerical viscous fluxes. We use the fourth-order $\alpha$-damping scheme of Chamarthi \cite{chamarthimig2022}, which is based on the $\alpha$-damping approach of Nishikawa \cite{nishikawa2011two}. Chamarthi et al. \cite{chamarthi2022importance} has shown that the $\alpha$-damping approach prevents odd-even decoupling and also plays an important role in turbulent flow simulations \cite{chamarthi2023role}. For simplicity and without loss of generality, we consider a one-dimensional scenario. The cell center numerical viscous flux derivative is:

\begin{equation}
    \left. \frac{\text{d} \check{\mathbf{F}}^v}{\text{d} x} \right|_{i} = \frac{1}{\Delta x} \left( \check{\mathbf{F}}^{v}_{i+\frac{1}{2}} - \check{\mathbf{F}}^{v}_{i-\frac{1}{2}} \right),
\end{equation}

\noindent The cell interface numerical viscous flux is:

\begin{equation}
    \check{\mathbf{F}}^v_{i+\frac{1}{2}} = 
    \begin{pmatrix}
        0 \\
        -\tau_{i+\frac{1}{2}} \\
        -\tau_{i+\frac{1}{2}} u_{i+\frac{1}{2}} + q_{i+\frac{1}{2}} \\
    \end{pmatrix},
\end{equation}

\noindent where,

\begin{subequations}
    \begin{gather}
        \tau_{i+\frac{1}{2}} = \frac{4}{3} \hat{\mu}_{i+\frac{1}{2}} \left. \frac{\partial u}{\partial x} \right|_{i+\frac{1}{2}},
        \quad
        q_{i+\frac{1}{2}} = -\hat{\kappa}_{i+\frac{1}{2}} \left. \frac{\partial T}{\partial x} \right|_{i+\frac{1}{2}}.
        \tag{\theequation a--\theequation b}
        \label{eqn:interfaceViscousStress}
    \end{gather}
\end{subequations}

\noindent For an arbitrary variable $\phi$, the $\alpha$-damping approach computes cell interface gradients as:

\begin{equation}
    \left. \frac{\partial \phi}{\partial x} \right|_{i+ \frac{1}{2}} = \frac{1}{2} \left( \left. \frac{\partial \phi}{\partial x} \right|_{i} + \left. \frac{\partial \phi}{\partial x} \right|_{i+1} \right) + \frac{\alpha}{2 \Delta x} \left( \phi_R - \phi_L \right), 
\end{equation}

\noindent where $\alpha = 4$ and, 

\begin{subequations}
    \begin{gather}
        \phi_L = \phi_i + \left. \frac{\partial \phi}{\partial x} \right|_{i} \frac{\Delta x}{2},
        \quad
        \phi_R = \phi_{i+1} - \left. \frac{\partial \phi}{\partial x} \right|_{i+1} \frac{\Delta x}{2},
        \tag{\theequation a--\theequation b}
    \end{gather}
\end{subequations}

\noindent The gradients at cell centers are the same ones computed in Equation (\ref{eqn:firstDerivative}). Since the conservative variables are used for inviscid flux interpolation, the necessary velocity gradients for the viscous fluxes are calculated from the following relations:

\begin{subequations}
    \begin{gather}
        \frac{\partial u}{\partial x} = \frac{1}{{\rho}^{2}} \left( {\rho} \frac{\partial \rho u}{\partial x} - {\rho} {u} \frac{\partial \rho }{\partial x} \right),
        \quad
        \frac{\partial v}{\partial x} = \frac{1}{{\rho}^{2}} \left( {\rho} \frac{\partial \rho v}{\partial x} - {\rho} {v} \frac{\partial \rho}{\partial x} \right).
        \tag{\theequation a--\theequation b}
        \label{eqn:compute_gradients}
    \end{gather}
\end{subequations}

\begin{figure}[H]
    \centering
    \includegraphics[width=0.5\textwidth]{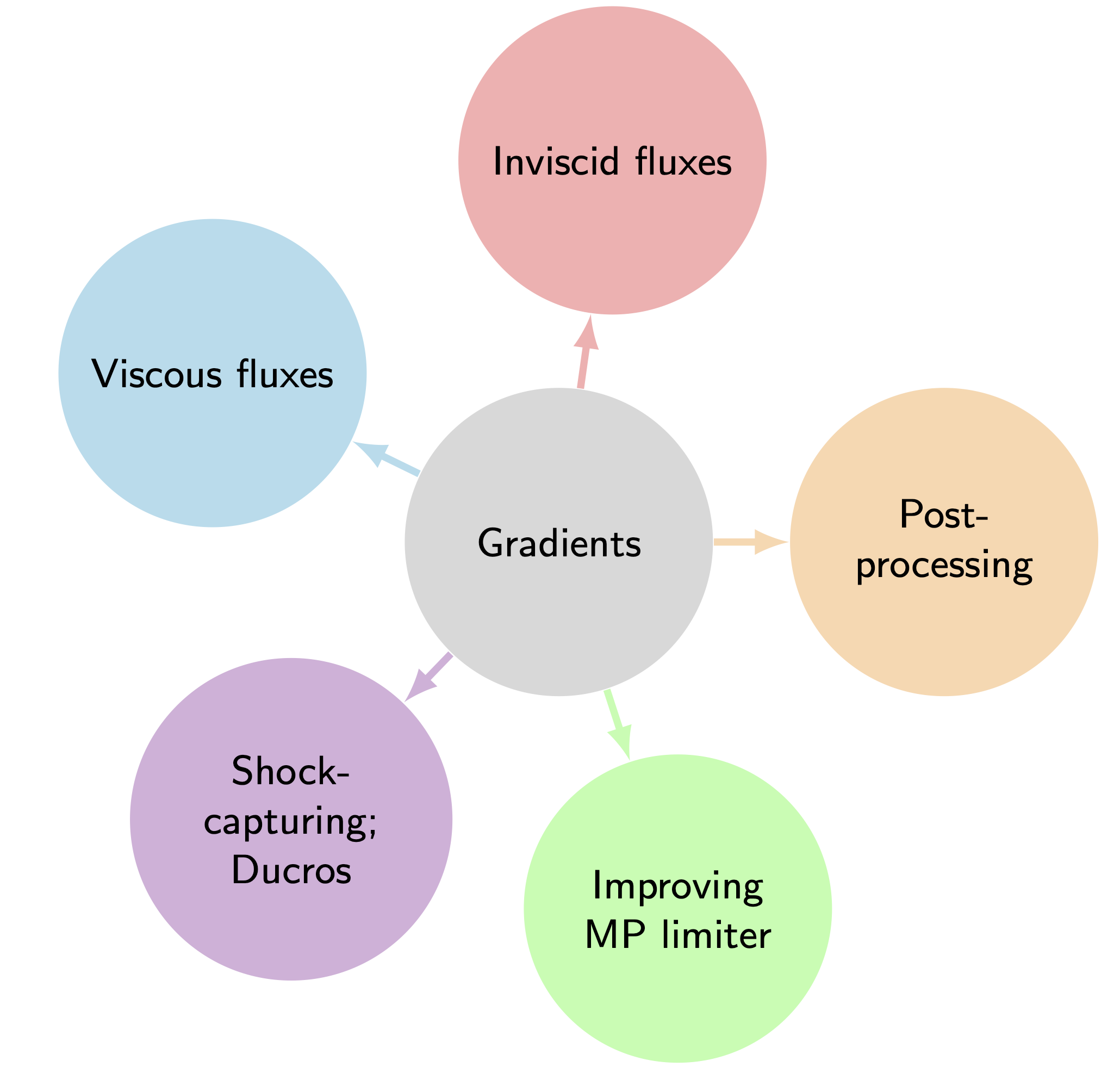}
    \caption{Locations where gradients are re-used in the GBR method.}
    \label{fig:grad}
\end{figure}

Fig. \ref{fig:grad} summarizes the various subroutines where gradients are re-used. Once computed, the gradients are used for the inviscid fluxes (Equation (\ref{eqn:legendre})), for the viscous fluxes (Equation (\ref{eqn:interfaceViscousStress})), for the shock detector (Equation (\ref{eqn:ducros})), to improve the MP limiter characteristics (Equation (\ref{mp_improve})) as discussed in Chamarthi \cite{chamarthimig2022}, and for post-processing quantities such as enstrophy or Q-criterion.

\subsection{Time Integration}

The explicit third-order total-variation-diminishing Runge-Kutta (RK3TVD) \cite{gottlieb1998total} method \textcolor{black}{was} used for time integration. The timestep, $\Delta t$, was computed from the CFL condition. We used both a convective and viscous analogue of the CFL condition. For all simulations, $\mathrm{CFL} = 0.2$. The convective $\Delta t$ was computed from:

\begin{equation}
    \Delta t^c = \min \left( \frac{\Delta x}{\lvert u \rvert + c}, \frac{\Delta y}{\lvert v \rvert + c}, \frac{\Delta z}{\lvert w \rvert + c} \right).
\end{equation}

\noindent The viscous $\Delta t$ was computed from:

\begin{equation}
    \Delta t^v = \frac{1}{\alpha} \min \left( \frac{\Delta x^2}{\hat{\nu}}, \frac{\Delta y^2}{\hat{\nu}}, \frac{\Delta z^2}{\hat{\nu}} \right),
\end{equation}

\noindent where $\alpha = 4$ corresponds to that employed in the viscous spatial discretization method and $\hat{\nu} = \hat{\mu}/\rho$ is the local, scaled kinematic viscosity \cite{chamarthi2023implicit,chamarthimig2022,chamarthi2022importance,Nishikawa2010}. Finally, the timestep was computed from:

\begin{equation}
    \Delta t = \mathrm{CFL} \times \min \left( \Delta t^c, \Delta t^v \right).
\end{equation}

\section{\label{sec:results}Numerical Results}

In this work, comparisons of the proposed method are made with the fifth-order TENO scheme \cite{fu2019low}. Only multi-dimensional flow simulations were carried out as the Ducros sensor is only valid for multi-dimensional flows. The example test cases are organized so that the proposed approach's benefits are clearly highlighted in a step-by-step manner.

\begin{example}\label{ex:rp}{Riemann Problem}
\end{example}

For the first example, we considered the two-dimensional Riemann problem of configuration 3 \cite{schulz1993numerical,fu2019low}. The initial conditions for the configuration \textcolor{black}{are}:

\begin{equation}
    (\rho, u, v, p)= 
    \begin{cases}
        (1.5, 0, 0, 1.5) & \text{if } x > 0.5, y > 0.5, 
        \\
        (0.5323, 1.206, 0, 0.3) & \text{if } x < 0.5, y > 0.5, 
        \\
        (0.138, 1.206, 1.206, 0.029) & \text{if } x < 0.5, y < 0.5, 
        \\
        (0.5323, 0, 1.206, 0.3) & \text{if } x > 0.5, y < 0.5. 
    \end{cases}
    \label{ex:rp1}
\end{equation}

\begin{figure}[H]
        \centering\offinterlineskip
        \subfigure[TENO]{\includegraphics[width=0.45\textwidth]{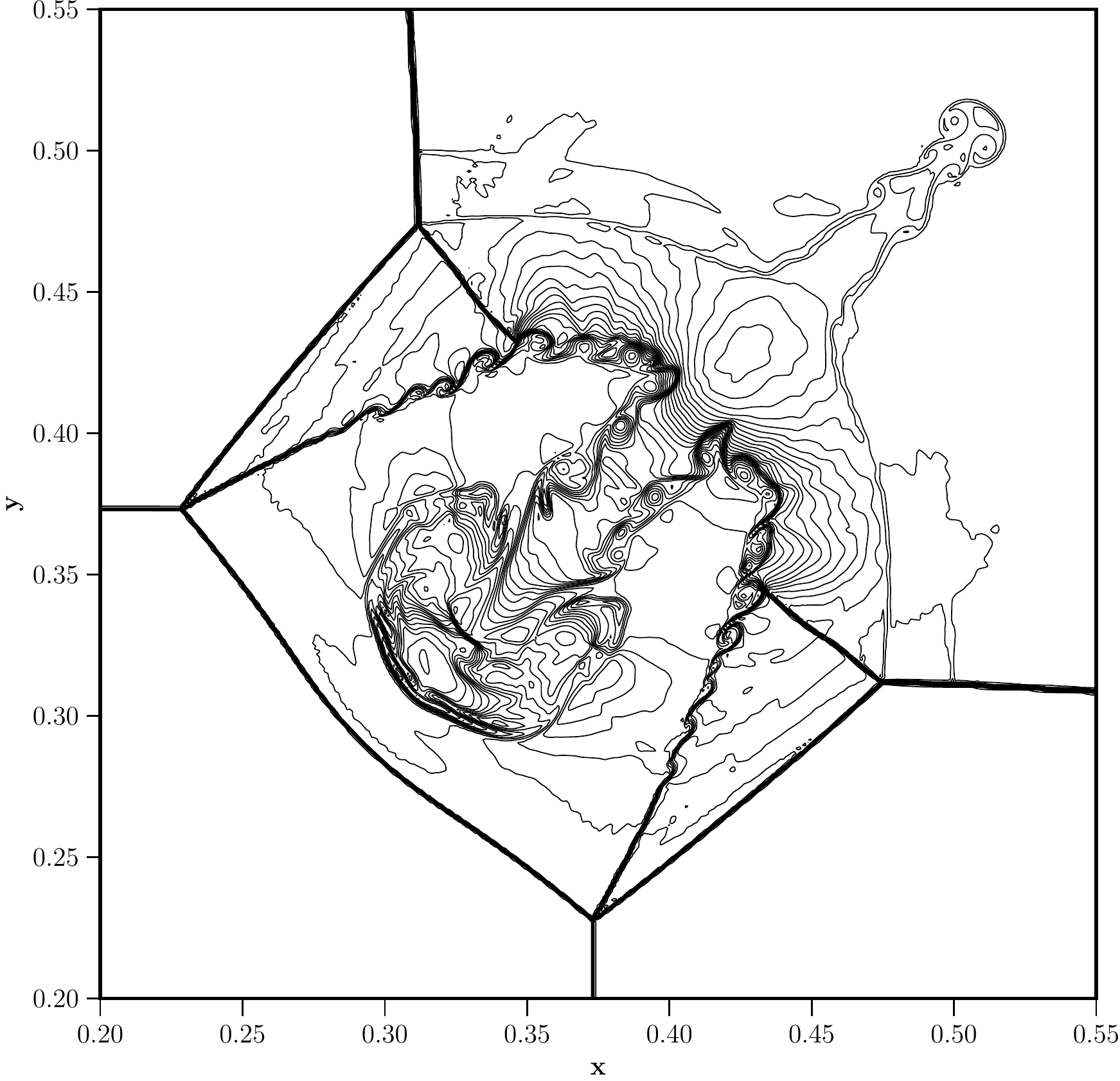}
        \label{fig:TENO_r12}}
        \subfigure[MIG-D]{\includegraphics[width=0.45\textwidth]{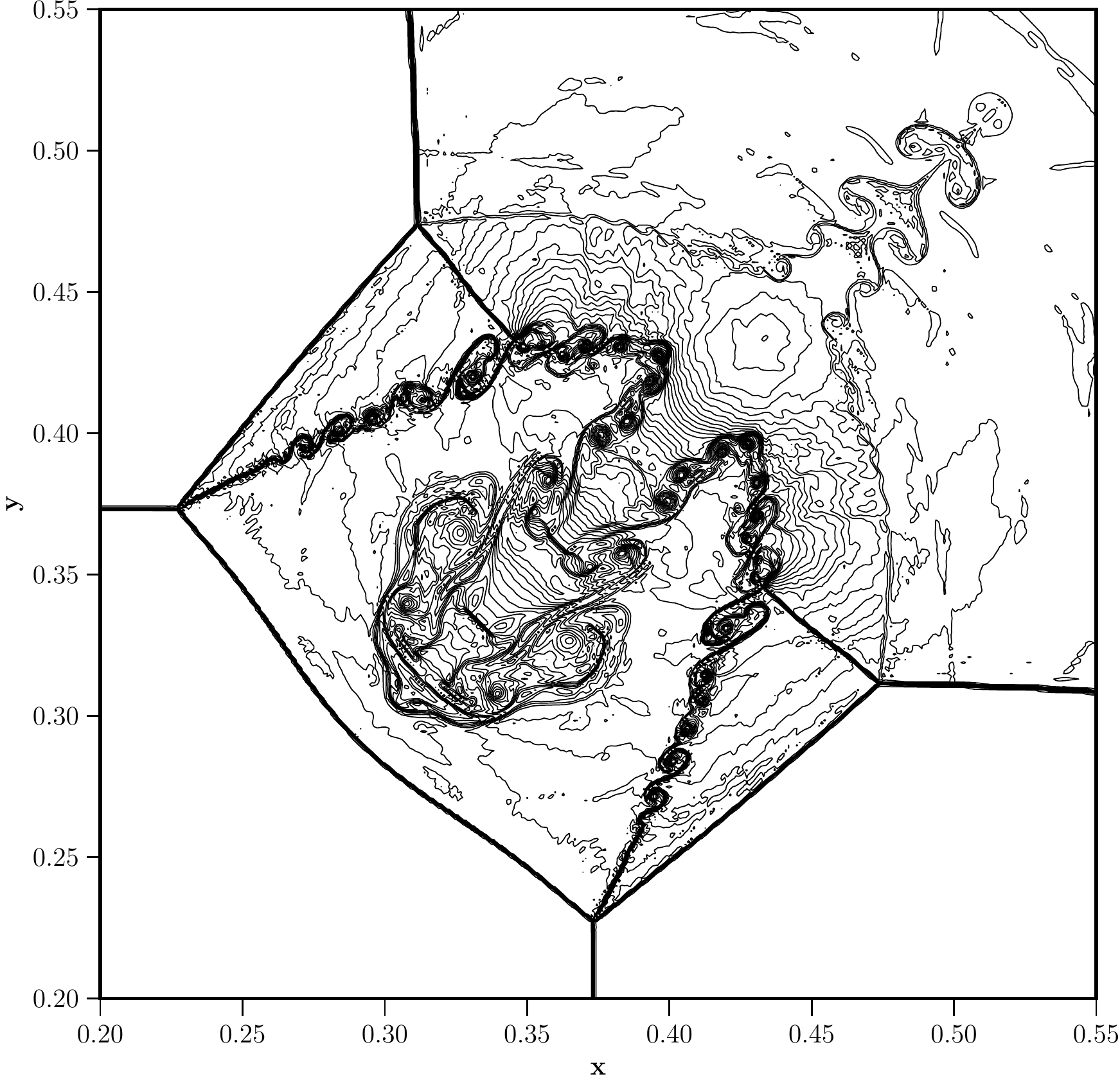}
        \label{fig:migd_r12}}
        \subfigure[MIG]{\includegraphics[width=0.45\textwidth]{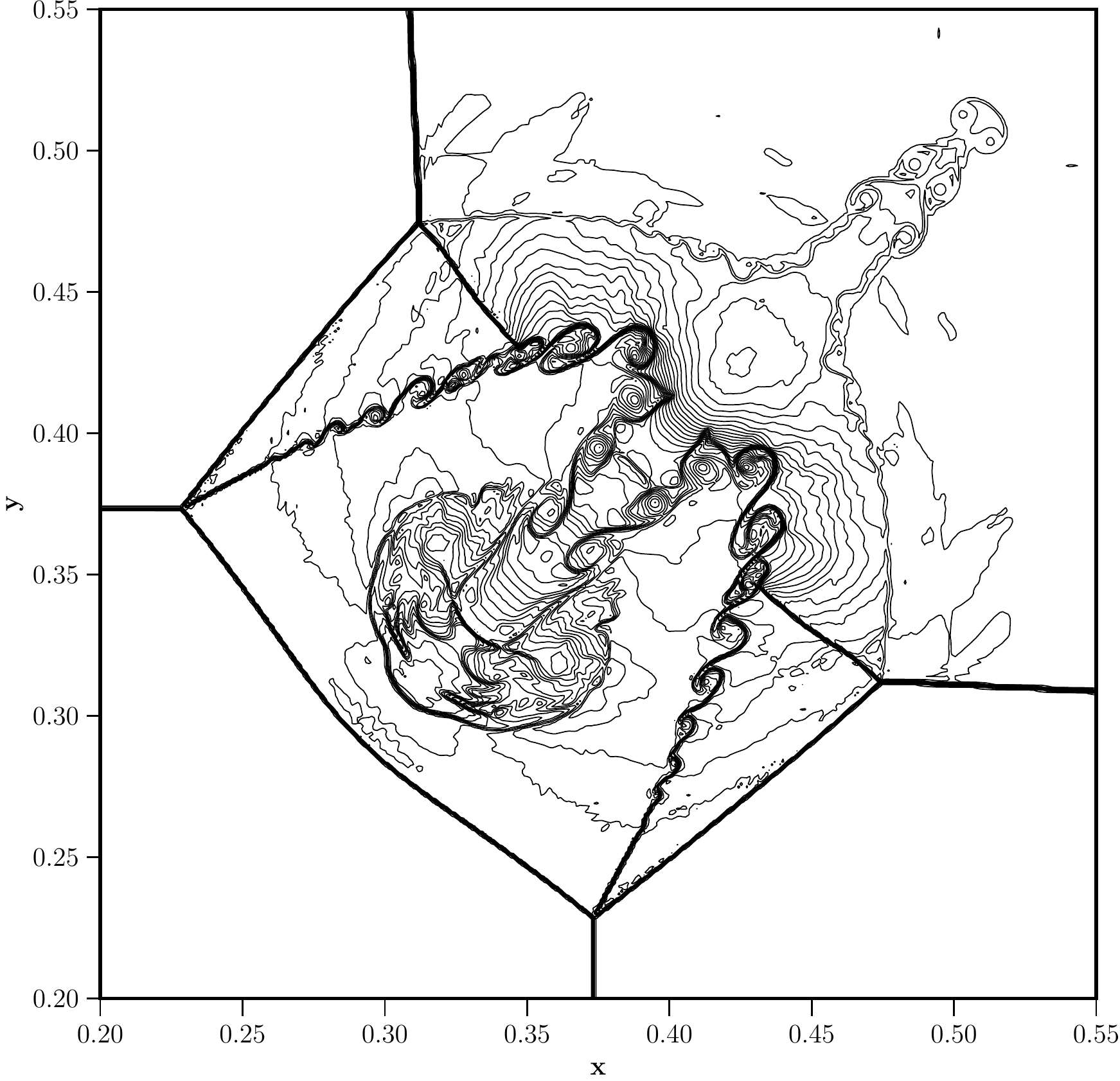}
        \label{fig:MIGb_r12}}
        \subfigure[MIG-S]{\includegraphics[width=0.45\textwidth]{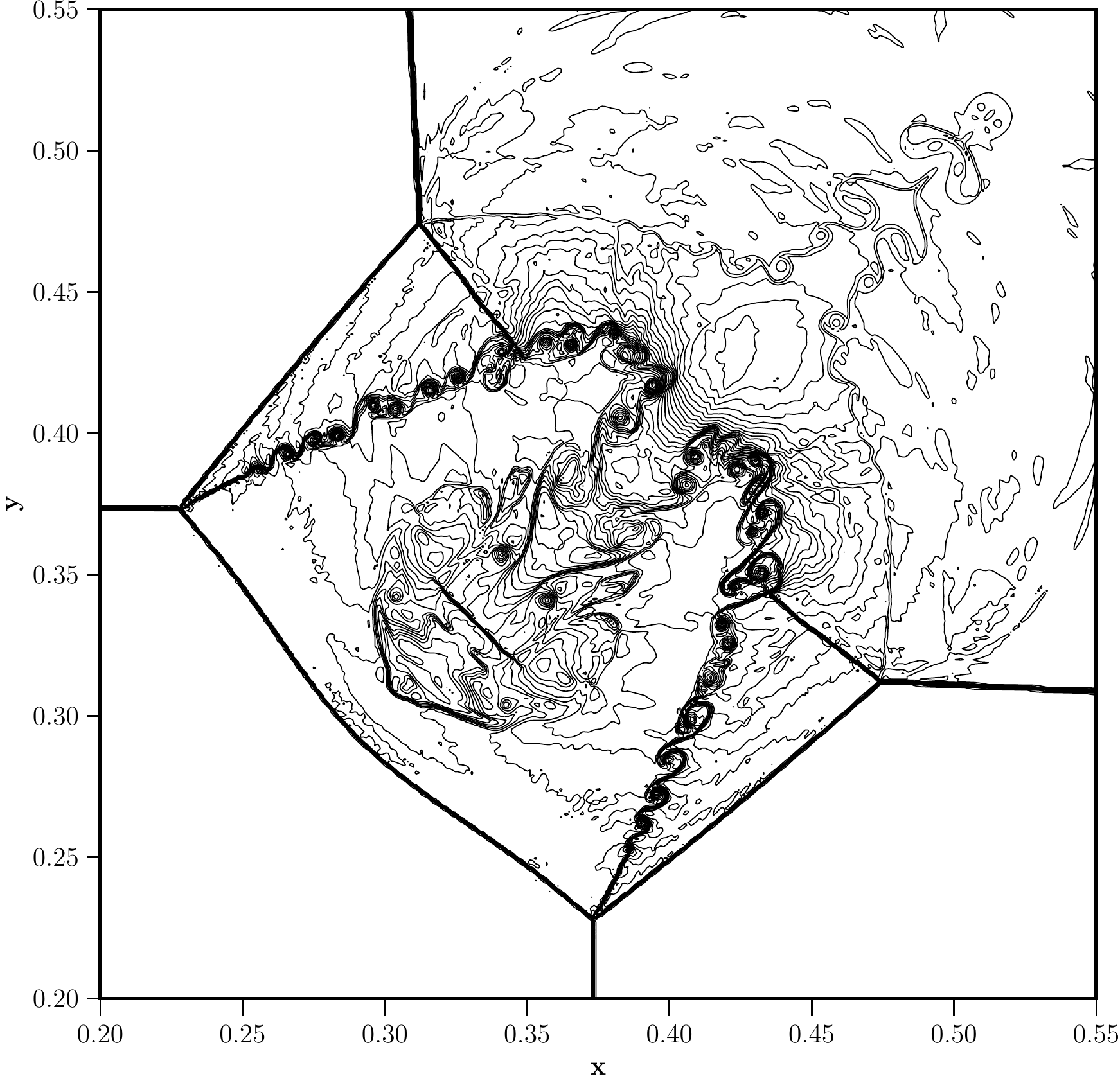}
        \label{fig:MIGs_r12}}
        \caption{Density contours of Example \ref{ex:rp} for the considered schemes. The figures are drawn with 42 contours. (a) TENO. (b) MIG-D. (c) MIG. (d) MIG-S.}
        \label{fig_riemann}
\end{figure}

\noindent The above initial conditions produce four shocks at the interfaces of the four quadrants. The Kelvin-Helmholtz instability along the slip lines form small-scale structures, which commonly serve as a benchmark for the numerical dissipation of a given scheme. The computational domain for this test case was $[x,y] = [0,1] \times [0,1]$ and simulations were carried out using a uniform grid of $1024 \times 1024$ until a final time, $t_f = 0.3$. Non-reflective boundary conditions were used at all domain boundaries. 

The density contours obtained by the considered schemes are presented in Figs \ref{fig_riemann}. The small-scale vortices produced by the TENO scheme (Fig. \ref{fig:TENO_r12}), are the most dissipated of all presented results, which is consistent with results shown in \cite{chamarthi2023implicit,chamarthimig2022}. Furthermore, the vortices resolved by the MIG-D scheme are \textcolor{black}{more pronounced than the MIG and MIG-S schemes. For this qualitative assessment, the number of vortices resolved and their resolution is indicative of a scheme's low-dissipative characteristics.} The discontinuity detection criterion used for shock detection is the only difference between the MIG-D, MIG-S, and MIG schemes. These results indicate that the MP limiting criterion, given by Equation (\ref{eqn:mpLimitingCriterion}), was activated far too frequently and dissipated the small-scale structures in comparison with the Ducros sensor, given by Equation (\ref{eqn:ducrosLimitingCriterion}).

It is important to note that this test case does not involve any contact discontinuities; therefore, the Ducros sensor performed without any difficulties. The selective discontinuity criterion used for the MIG-S scheme, given by Equations (\ref{eqn:newSensorCriterion}), did get activated in certain additional regions. As such, it produced slightly more dissipative results than the MIG-D scheme.

\begin{example}\label{ex:dmr}{Double Mach Reflection}
\end{example}

In the second example, the Double Mach Reflection (DMR) case of Woodward and Collela \cite{woodward1984numerical} was considered. A Mach $10$ unsteady planar shock wave impinges on a $30^{\circ}$ inclined surface and produces complex flow features. The computational domain for this test case was $[x,y] = [0,3] \times [0,1]$ and the simulations were carried out on a uniform grid of $768 \times 256$ with the following initial conditions:

\begin{equation}
    (\rho, u, v, p)= 
    \begin{cases}
        (1.4, 0, 0, 1) & \text{if } y < 1.732 \left( x - 0.1667 \right), 
        \\
        (8, 7.145, -4.125, 116.8333) & \text{otherwise}.
    \end{cases}
    \label{ex:dmr1}
\end{equation}

\begin{figure}[H]
        \centering\offinterlineskip
        \subfigure[TENO]{\includegraphics[width=0.42\textwidth]{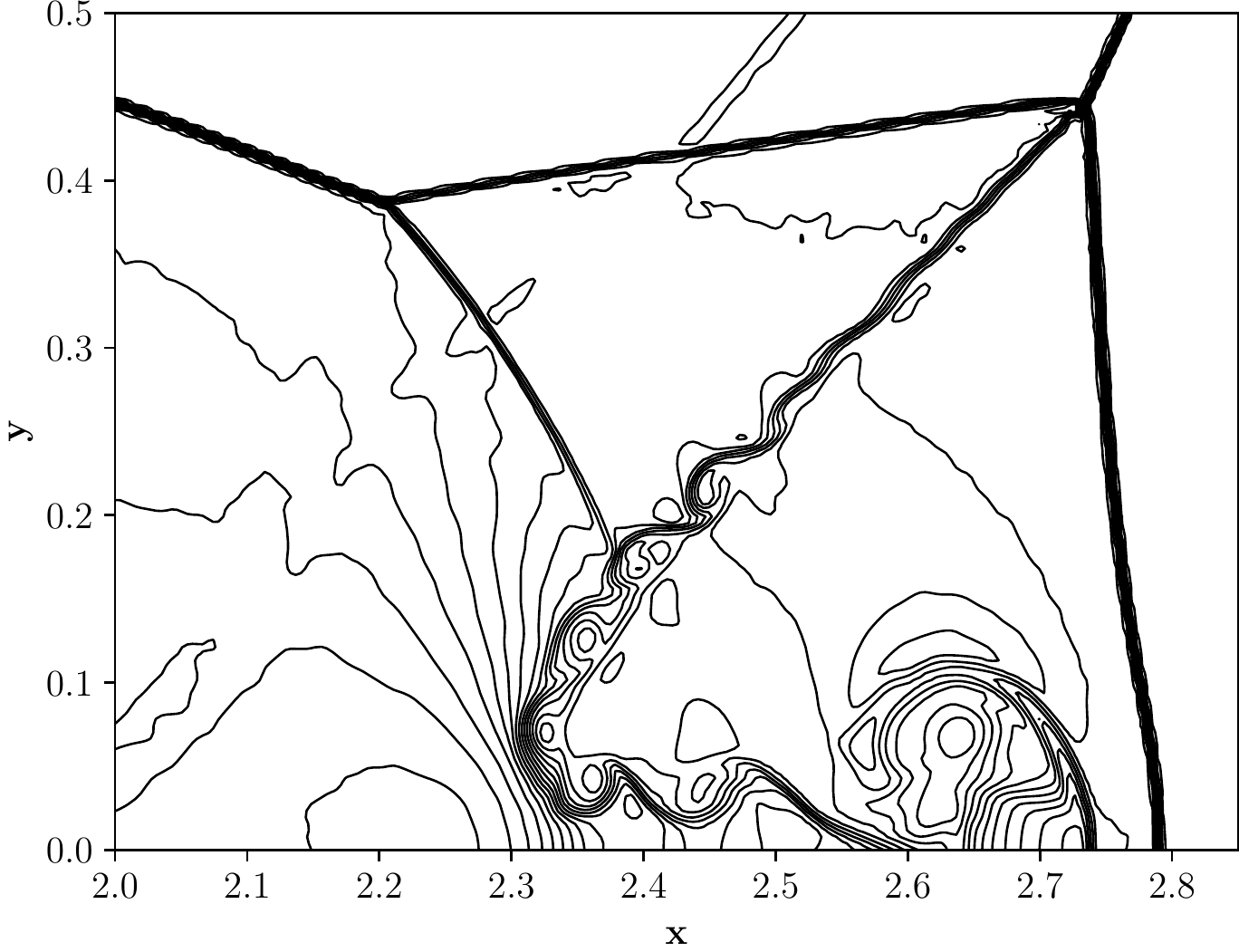}
        \label{fig:teno_dmr}}
        \subfigure[MIG-D]{\includegraphics[width=0.42\textwidth]{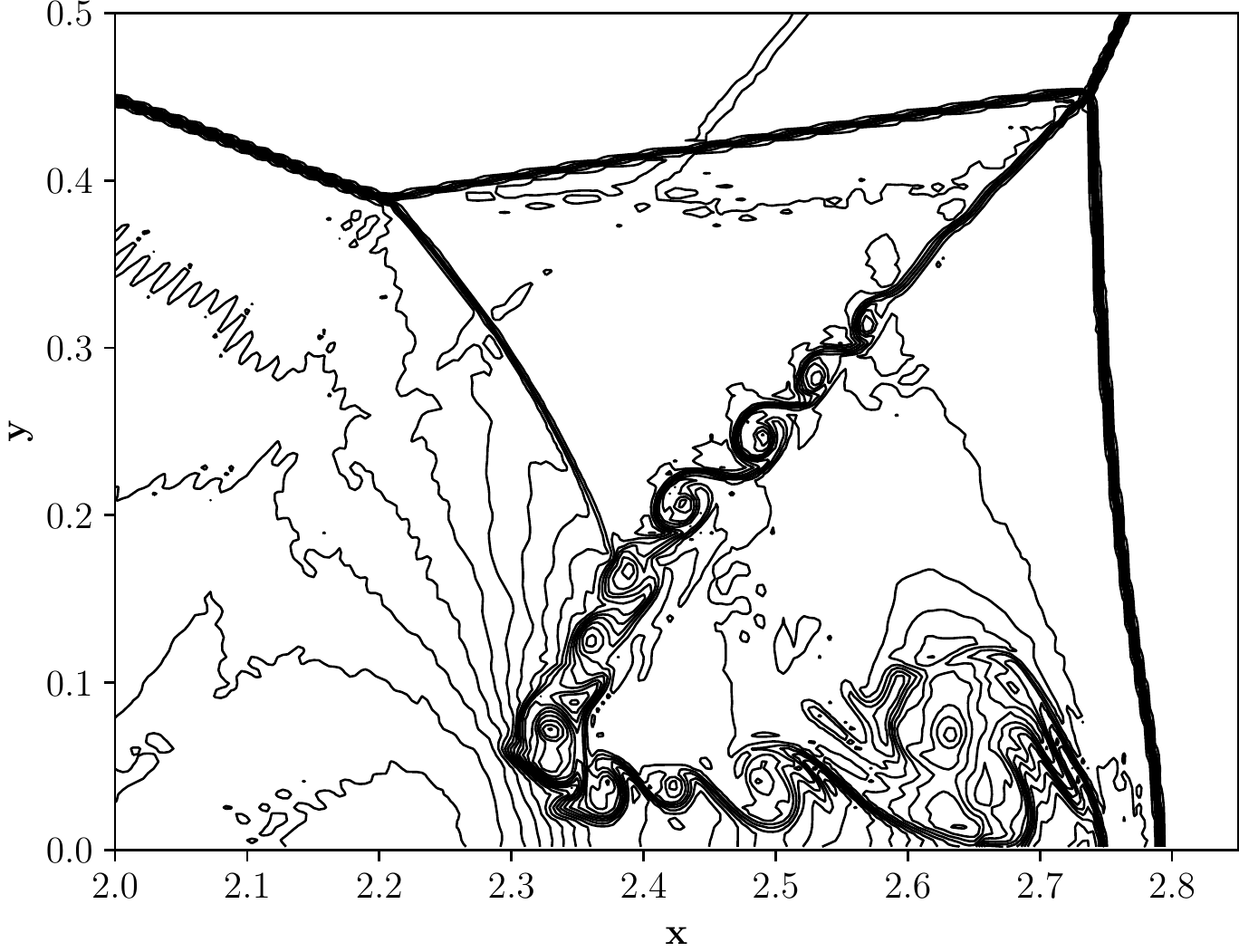}
        \label{fig:migd_dmr}}
        \subfigure[MIG]{\includegraphics[width=0.42\textwidth]{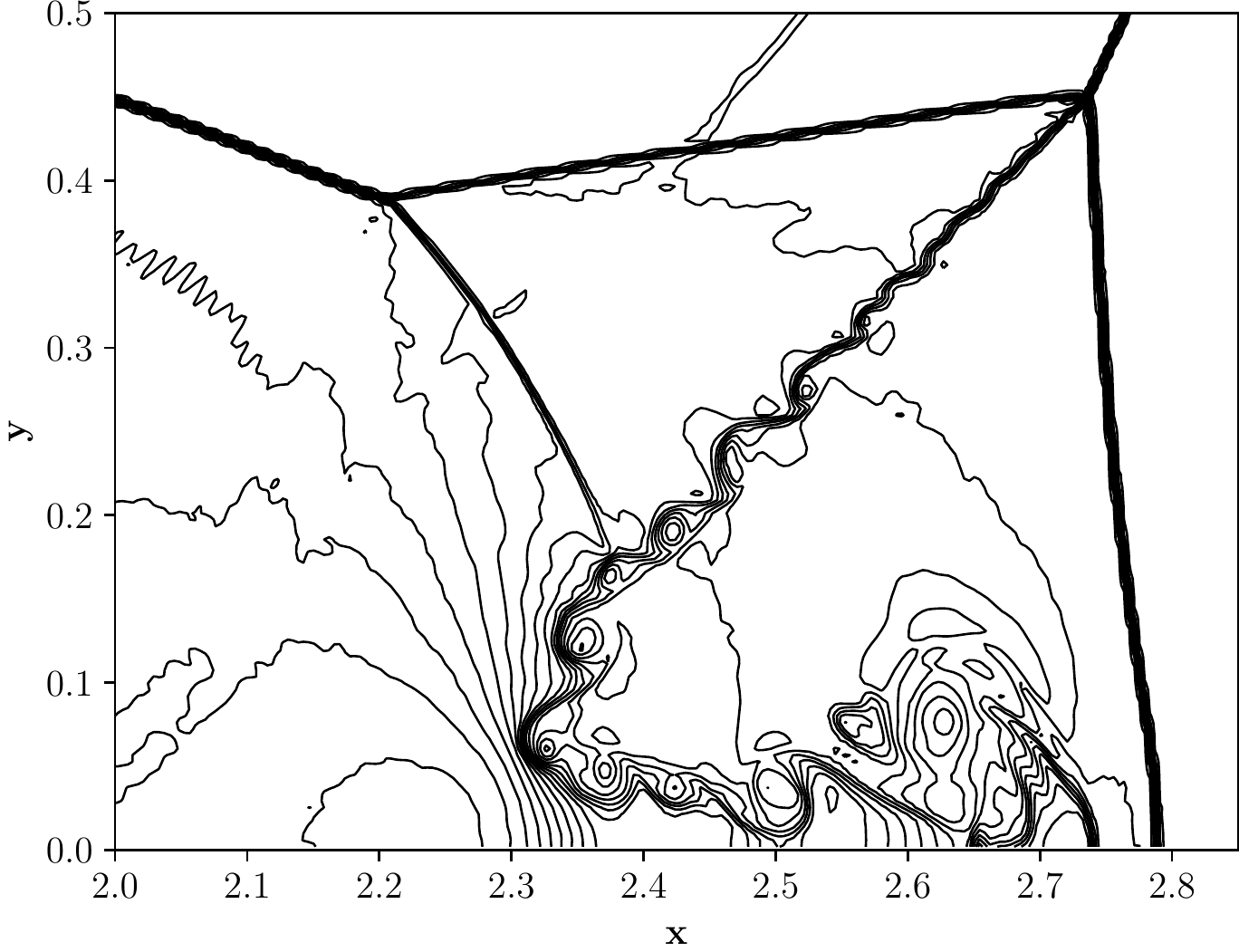}
        \label{fig:migb_dmr}}
        \subfigure[MIG-S]{\includegraphics[width=0.42\textwidth]{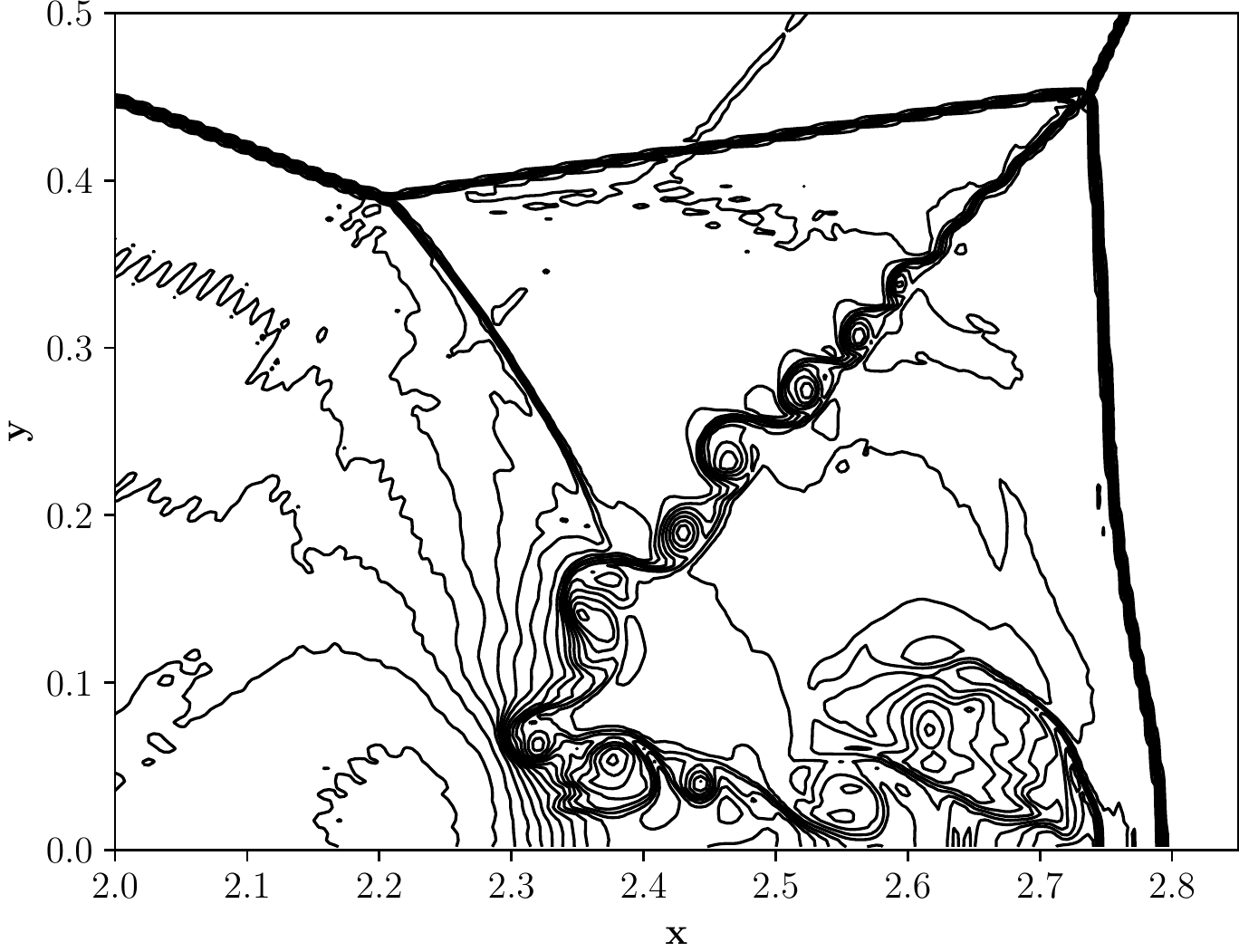}
        \label{fig:migs_dmr}}
        \caption{Density contours of the zoomed in Mach stem region of Example \ref{ex:dmr} for the considered schemes. The figures are drawn with 38 contours. (a) TENO. (b) MIG-D. (c) MIG. (d) MIG-S.}
        \label{fig_doublemach}
\end{figure}

\noindent For this test case, the shock stable HLLC approximate Riemann solver of Fleischmann et al. \cite{fleischmann2020shock} was used, as it is well-known that the standard HLLC scheme leads to the carbuncle phenomenon. The simulations were carried out until a final time, $t_f = 0.3$. The bottom boundary was treated with reflecting wall conditions for $0.1667 < x \leq 4.0$ and post-shock conditions for $0.0 \leq x \leq 0.1667$. The top boundary was set with the exact solution of the time-dependent oblique shock. The left and right boundary conditions may be found in the original reference.

The density contours obtained by all the considered schemes are shown in Figs. \ref{fig_doublemach}. The TENO scheme was the most dissipative of all considered methods as evidenced by it's smearing of small-scale vortices near the Mach stems. In the results obtained by the MIG family of schemes, the vortices and near wall jets were better-captured by the MIG-D and MIG-S schemes than the standard MIG scheme. One can draw similar conclusions as in Example \ref{ex:rp}: the MP detection criterion was detecting small flow features as discontinuities and was limiting them, whereas the Ducros sensor identified shock waves appropriately and therefore, the overall scheme displayed low dissipation in other regions of the flow.

In the first two examples, the test cases did not feature a contact discontinuity. Therefore, the Ducros sensor performed without introducing oscillations. In the following test cases, we consider examples with a contact discontinuity which will show the Ducros sensor's deficiencies and also highlight the advantages of the proposed selective sensor.

\begin{example}\label{ex:rt}{Rayleigh-Taylor instability}
\end{example}

For the third example we considered the Rayleigh-Taylor instability. The Rayleigh-Taylor instability (RTI) is a hydrodynamic instability that occurs when a dense fluid is placed above a less dense fluid in a gravitational field. The instability arises due to an unstable density gradient, which causes the denser fluid to accelerate downwards and the lighter fluid to accelerate upwards, leading to complex mixing patterns between the two fluids. The objective of this test case is to show that the Ducros sensor leads to oscillations, which can be overcome by the selective sensor. The initial conditions of Rayleigh-Taylor instability were as follows \cite{xu2005anti}:

\begin{equation}
    (\rho, u, v, p)= 
    \begin{cases}
        (2.0, 0, -0.025 \sqrt{ 5p \cos(8 \pi x)/\left( 3 \rho \right)}, 2y + 1), & \text{if } 0 \leq y < 0.5,
        \\
        (1.0, 0, -0.025 \sqrt{ 5p \cos(8 \pi x)/\left( 3 \rho \right)}, y + 1.5),& \text{if } 0.5 \leq y \leq 1.
    \end{cases}
    \label{eu2D_RT}
\end{equation}

\begin{figure}[H]
        \centering\offinterlineskip
        \subfigure[TENO]{%
        \includegraphics[width=0.20\textwidth]{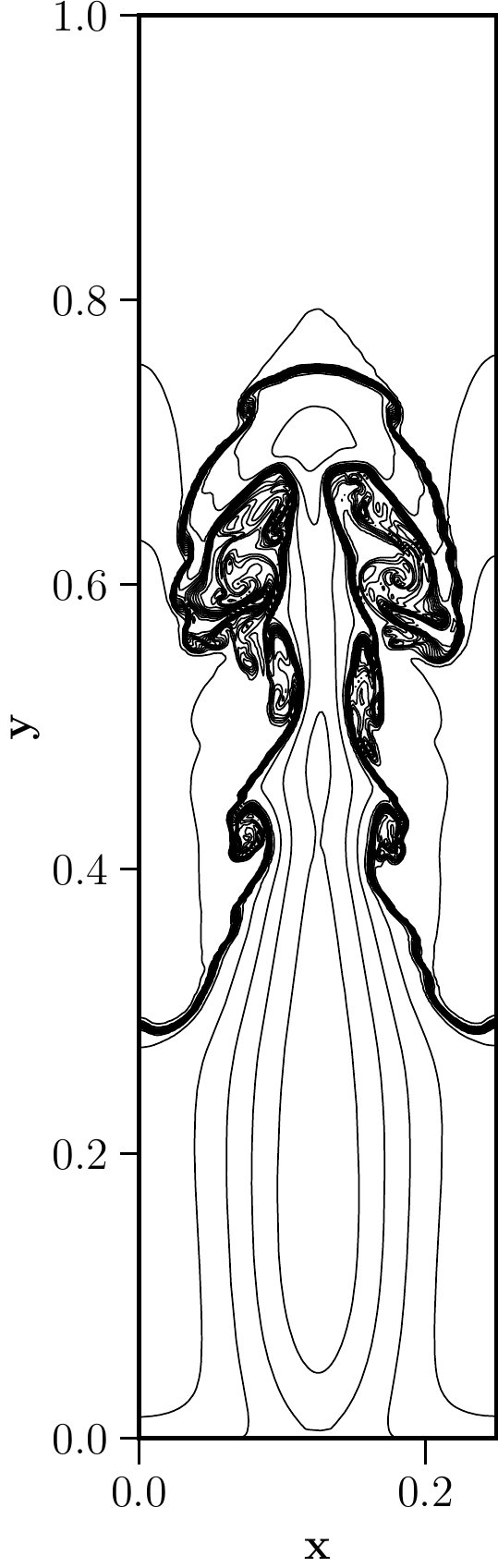}
        \label{fig:TENO_rt}}
        \subfigure[MIG-D]{%
        \includegraphics[width=0.20\textwidth]{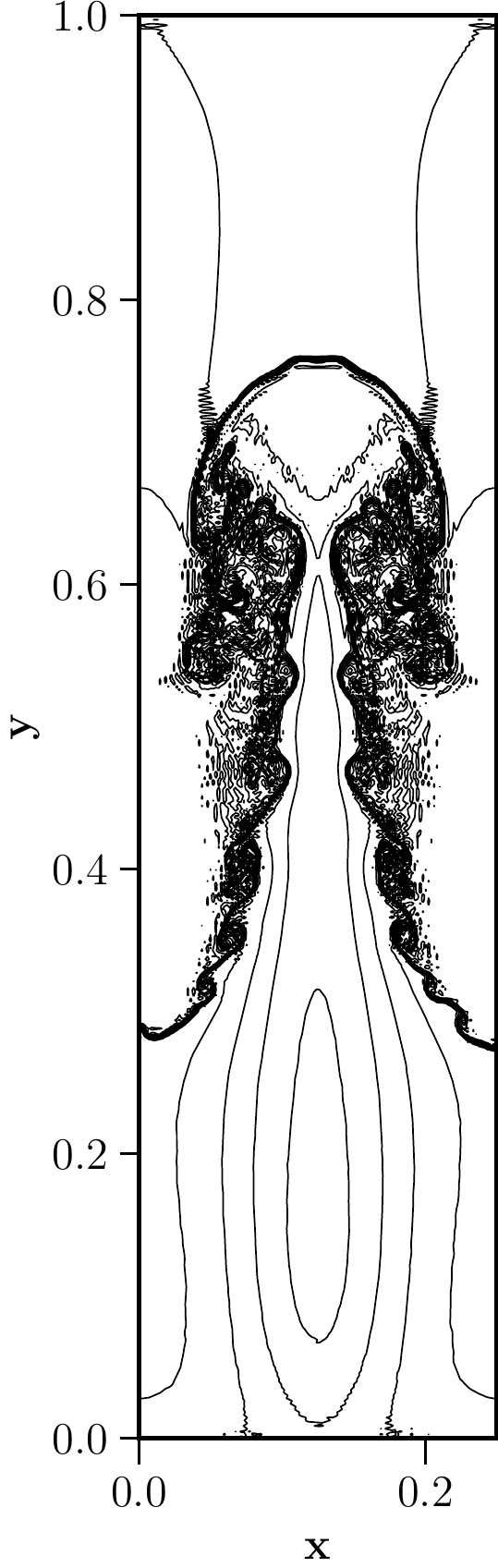}
        \label{fig:migd_rt}}
        \subfigure[MIG]{%
        \includegraphics[width=0.20\textwidth]{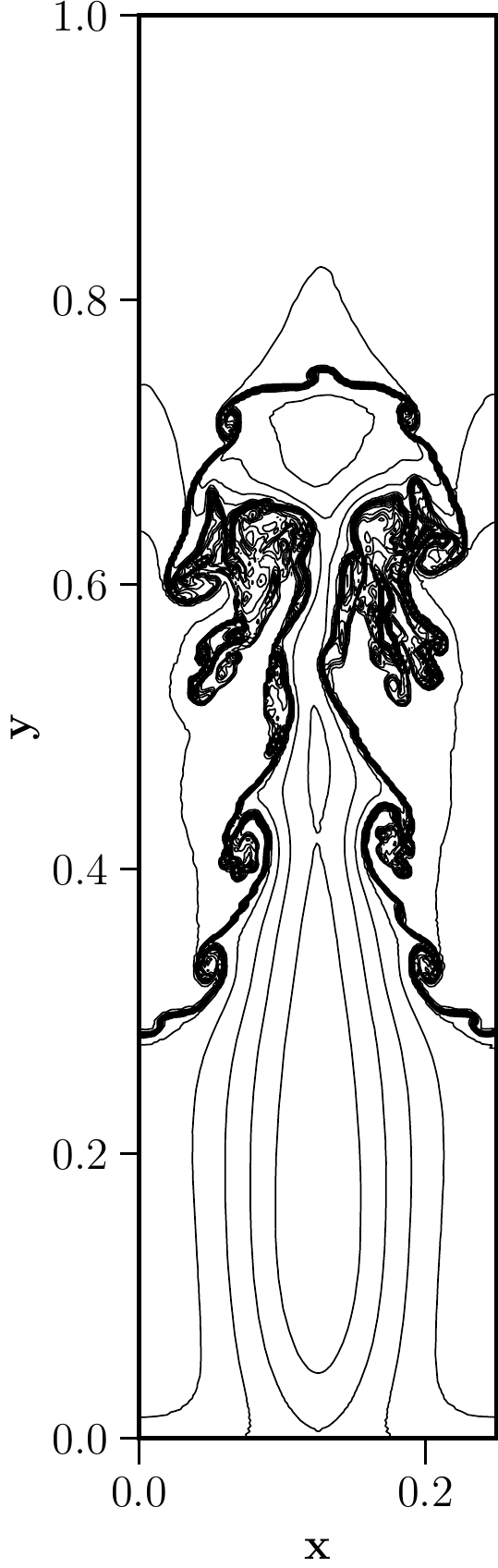}
        \label{fig:migb_rt}}
        \subfigure[MIG-S]{%
        \includegraphics[width=0.20\textwidth]{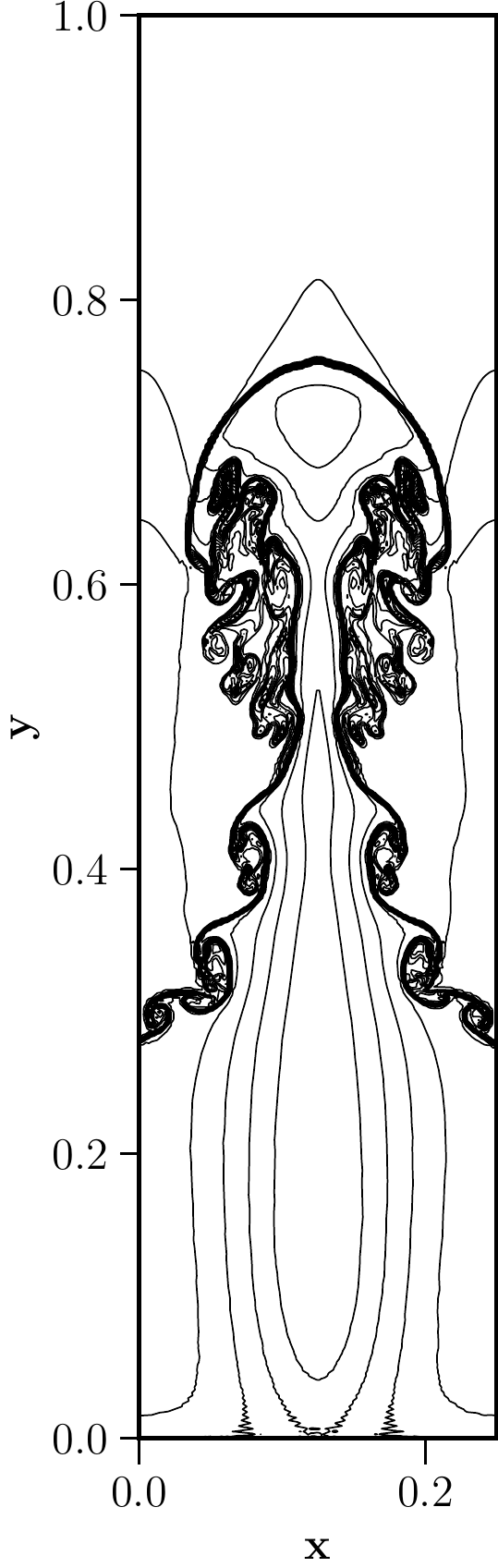}
        \label{fig:migs_rt}}
        \caption{Density contours of Example \ref{ex:rt} for the considered schemes. The figures are drawn with 20 contours. (a) TENO. (b) MIG-D. (c) MIG. (d) MIG-S.}
        \label{fig:2d-RT}
\end{figure}

The computational domain for this test case was $[x,y] = [0,1/4] \times [0,1]$  with $\gamma = 1.4$. Simulations were carried out on a uniform mesh of $128 \times 512$ until $t_f = 1.95$. The flow conditions were set to $\left( \rho, u, v, p \right) = \left( 1, 0, 0, 2.5 \right)$ on the top boundary and $\left( \rho, u, v, p \right) = \left( 2, 0, 0, 1 \right)$ on the bottom boundary. The respective source terms $S = (0, 0, \rho, \rho v)$ were added to the right-hand-side of the Euler equations. Figs. \ref{fig:2d-RT} display the density iso-contours of the considered schemes. It can be seen that the MIG-D scheme showed significant oscillations since the Ducros sensor cannot detect contact discontinuities. Conversely, the MIG-S scheme with the selective discontinuity sensor was free of oscillations. Furthermore, the MIG-S scheme significantly improved the resolution of the contact discontinuity and the small-scale finger-like features compared to the MIG and TENO schemes, \textcolor{black}{which qualitatively portrays it's superiority over them}. \\

\begin{example}\label{ex:rm} {Richtmyer--Meshkov instability}
\end{example}

For the fourth example, the two-dimensional single-mode Richtmyer-Meshkov instability (RMI) problem \cite{chamarthimig2022,terashima2009front} with the following initial conditions was considered:

\begin{equation}
    (\rho, u, v, p)= 
    \begin{cases}
        (5.04, 0, 0, 1), & \text{if } x < 2.9 - 0.1 \sin \left[ 2 \pi \left( y + 0.25 \right) \right],
        \\
        (1, 0, 0, 1),& \text{if } x < 3.2,
        \\
        (1.4112, -665/1556, 0, 1.628),& \text{otherwise}.
    \end{cases}
\end{equation}

The RMI is a type of hydrodynamic instability that arises when a shock wave interacts with a perturbed interface between two fluids of different densities. The instability leads to the formation of complex flow features including spikes, bubbles, and mixing layers that can significantly enhance the mixing between the two fluids. The computational domain for this test case was $[x,y] = [0,4] \times [0,1]$ and the simulation was conducted until $t_f = 9$, on a uniform mesh of $320 \times 80$. The boundary conditions can be found in Chamarthi \cite{chamarthimig2022}.

\begin{figure}[H]
        \centering\offinterlineskip
        \subfigure[TENO]{\includegraphics[width=0.25\textheight]{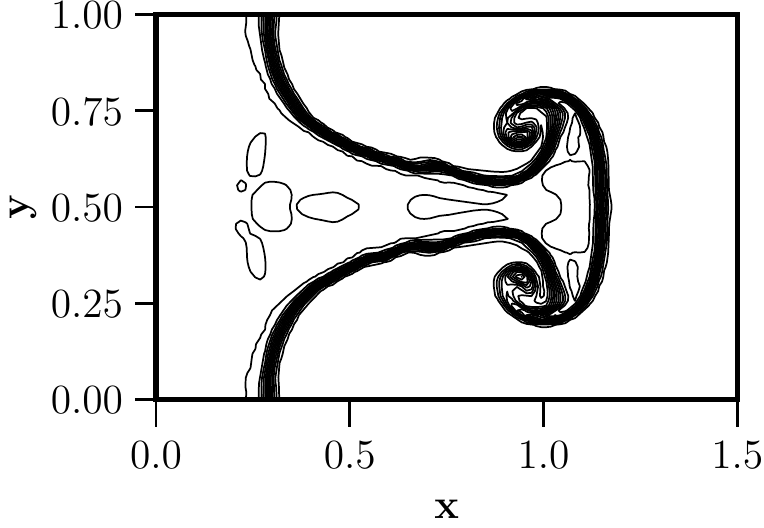}
        \label{fig:TENO_rm}}
        \subfigure[MIG-D, $t$ = 7]{\includegraphics[width=0.25\textheight]{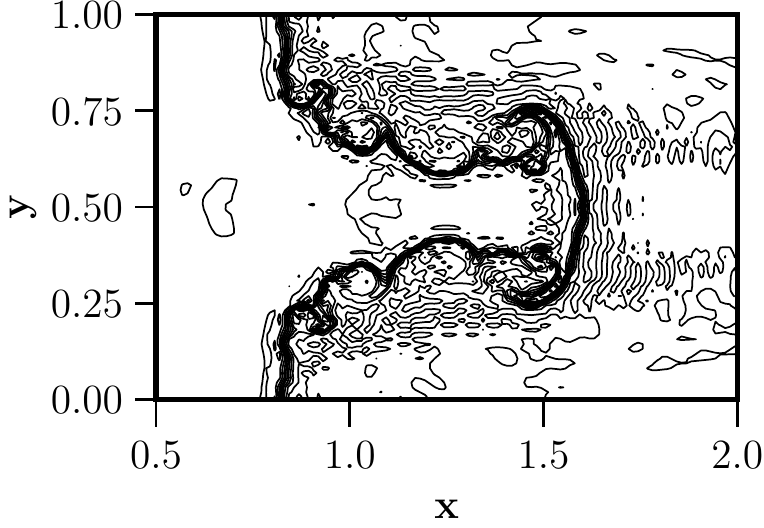}
        \label{fig:migd_rm}}
        \subfigure[MIG]{\includegraphics[width=0.25\textheight]{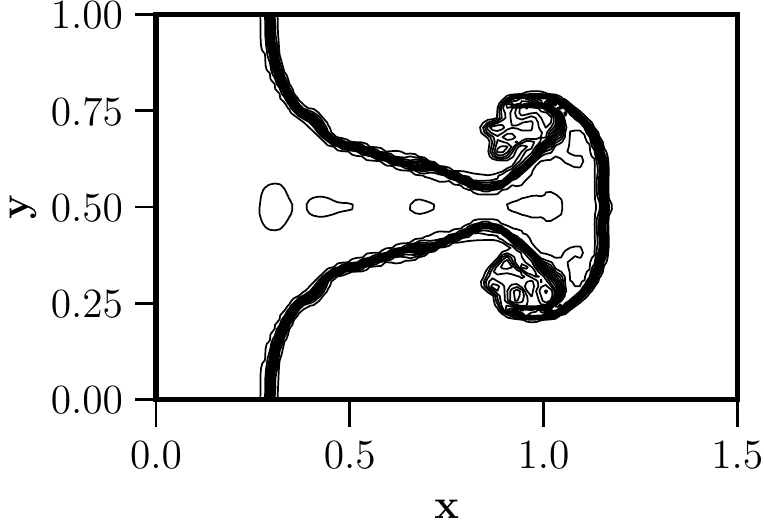}
        \label{fig:migb_rm}}
        \subfigure[MIG-S]{\includegraphics[width=0.25\textheight]{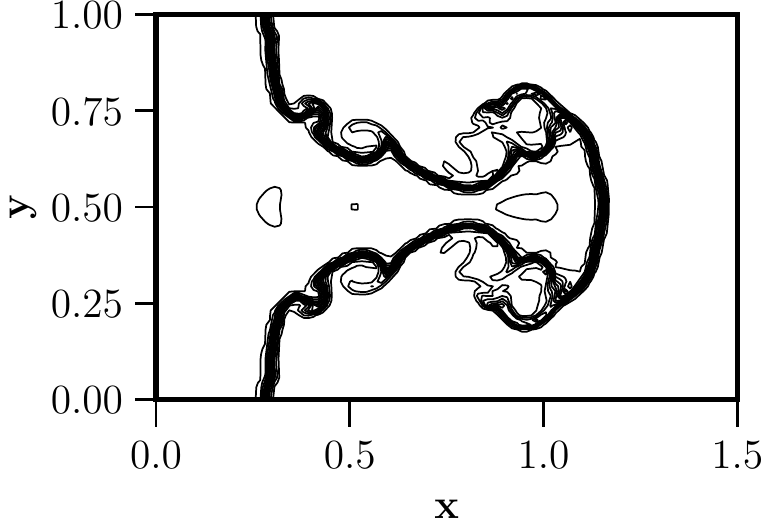}
        \label{fig:migs_rm}}
        \caption{Density contours of Example \ref{ex:rm} for the considered schemes. The figures are drawn with 20 contours. (a) TENO. (b) MIG-D. (c) MIG. (d) MIG-S.}
        \label{fig_RM}
\end{figure}

The density iso-contours computed using the considered schemes are shown in Fig. \ref{fig_RM}. Observing Fig. \ref{fig:migs_rm}, there is less numerical dissipation and better resolved small-scale roll-up vortices compared to the other schemes. Moreover, the material interface is thinner than in the TENO results. Sidharth and Candler \cite{gs2015large} found that in flows with RMI, the Ducros sensor generates spurious small scales (possibly oscillations) and breaks symmetry. The same phenomenon was observed in the MIG-D simulation, as seen in Fig. \ref{fig:migd_rm}. It is evident that the Ducros sensor cannot detect contact discontinuities and therefore leads to oscillations. On the other hand, the MIG-S approach does not show any oscillations as the selective detector approach can detect contact discontinuities appropriately. It is important to note that the MIG-D scheme failed to complete due to numerical oscillations until the prescribed final time of $t_f = 9$. As such, the results obtained at $t = 7$ are shown in Fig. \ref{fig:migd_rm}. 

\begin{example}\label{ex:kh}{Kelvin Helmholtz instability}
\end{example}

For the fifth example, we considered the Kelvin-Helmholtz instability (KHI). The KHI is another type of hydrodynamic instability that occurs when there is a velocity gradient between two fluids of different densities. The instability arises due to the unstable velocity gradient at the interface between the two fluids, leading to vortices and complex mixing patterns. It plays an important role in the evolution of the mixing layer and the transition to turbulence in three-dimensional turbulent cases. The test case has the following initial conditions:

\begin{equation}
    \begin{aligned}
        & \rho (x,y) = 
        \begin{cases}
            2, & \text{if } 0.25 < y \leq 0.75,
            \\
            1, & \text{otherwise},
        \end{cases}
        \\
        & u (x,y) = 
        \begin{cases}
            0.5, & \text{if } 0.25 < y \leq 0.75,
            \\
            -0.5, & \text{otherwise},
        \end{cases}
        \\
        & v (x,y)= 0.1 \sin (4 \pi x) \left\{ \exp \left[ -\frac{(y - 0.75)^{2}}{2 \sigma^{2}} \right] + \exp \left[ -\frac{(y - 0.25)^{2}}{2 \sigma^{2}} \right] \right\}, \quad \text{where } \sigma = 0.05/\sqrt {2}    
        \\
        & p(x,y) = 2.5
    \end{aligned}
\end{equation}

\begin{figure}[H]
        \centering\offinterlineskip
        \subfigure[TENO]{\includegraphics[width=0.3\textheight]{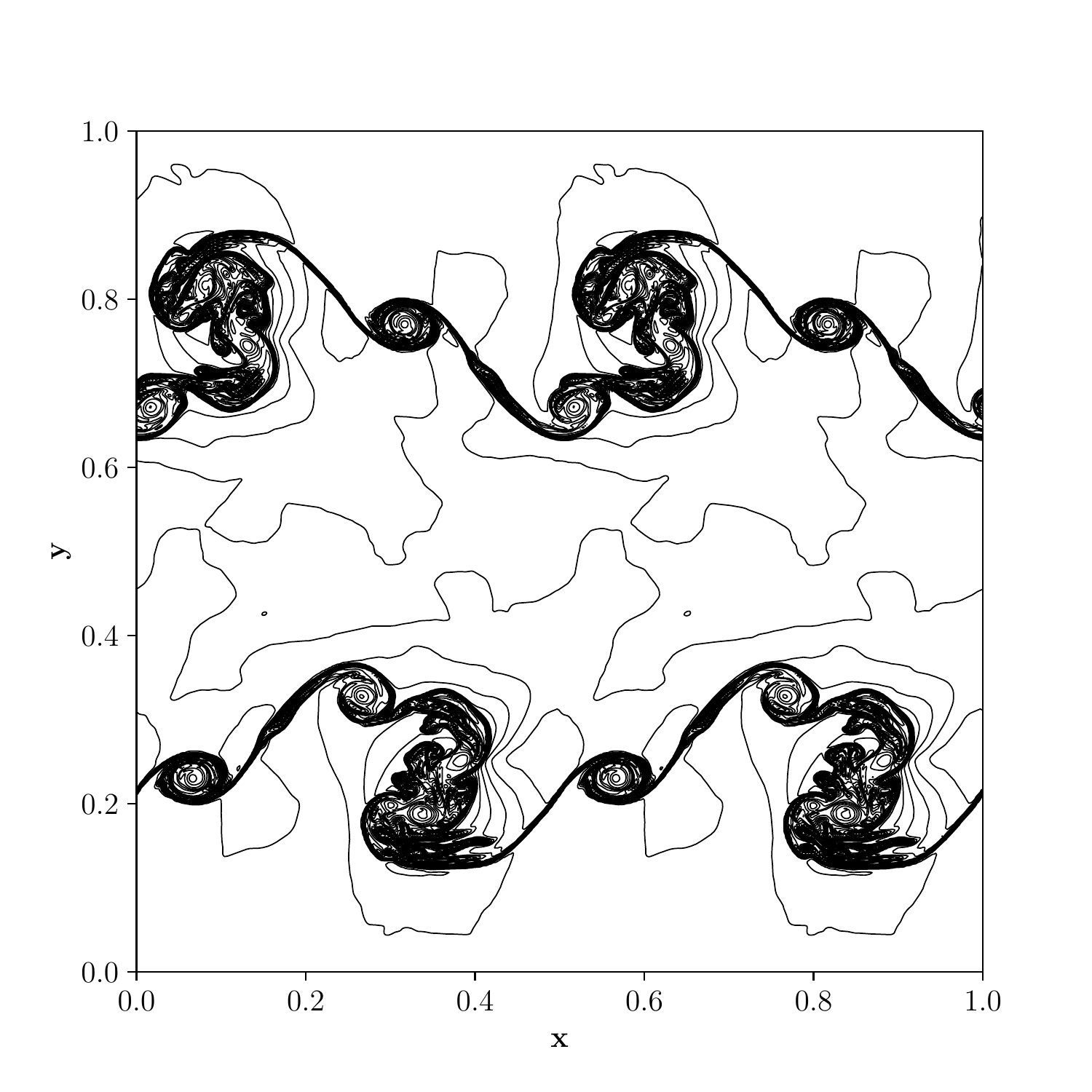}
        \label{fig:TENO_khi}}
        \subfigure[MIG-D]{\includegraphics[width=0.3\textheight]{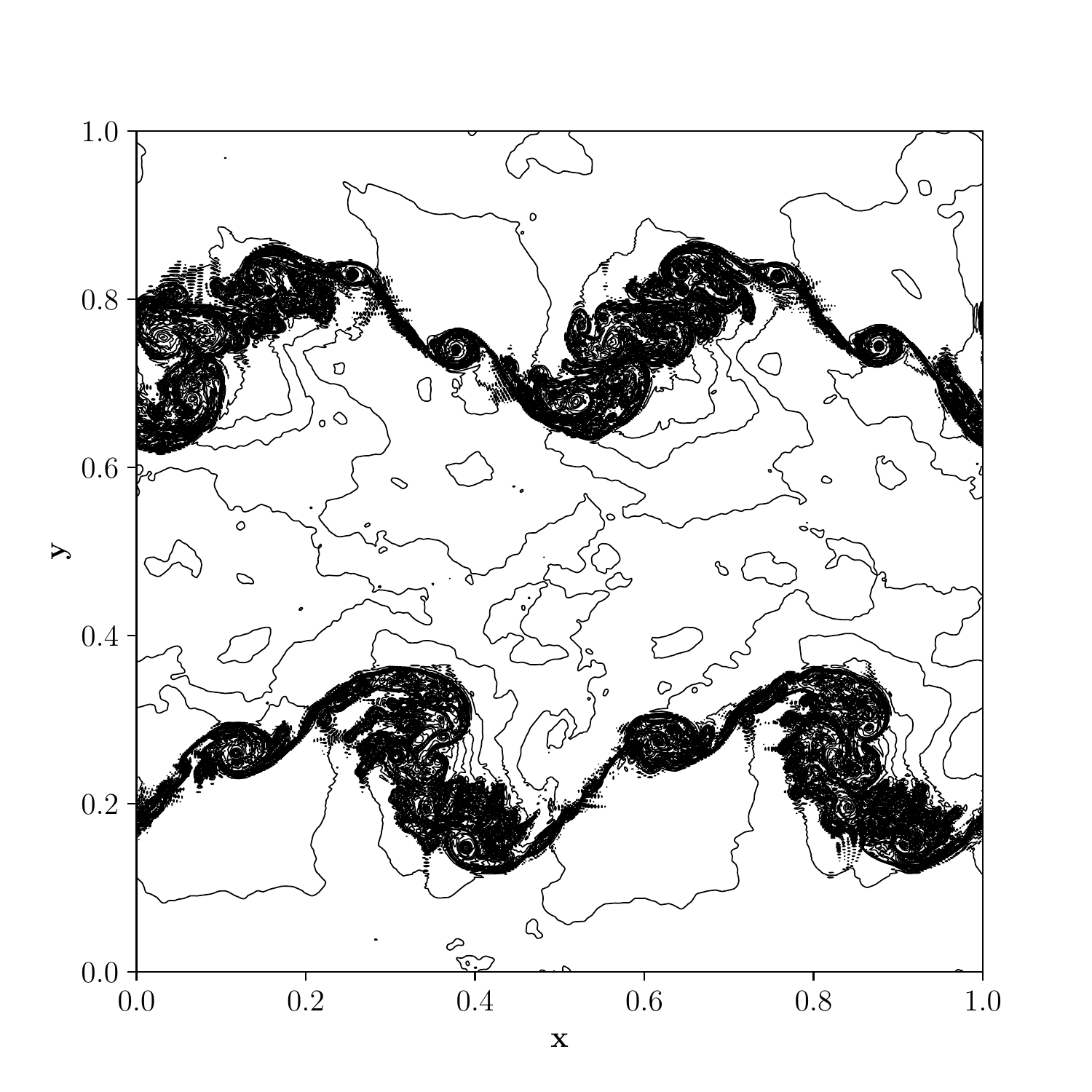}
        \label{fig:migd_khi}}
        \subfigure[MIG]{\includegraphics[width=0.3\textheight]{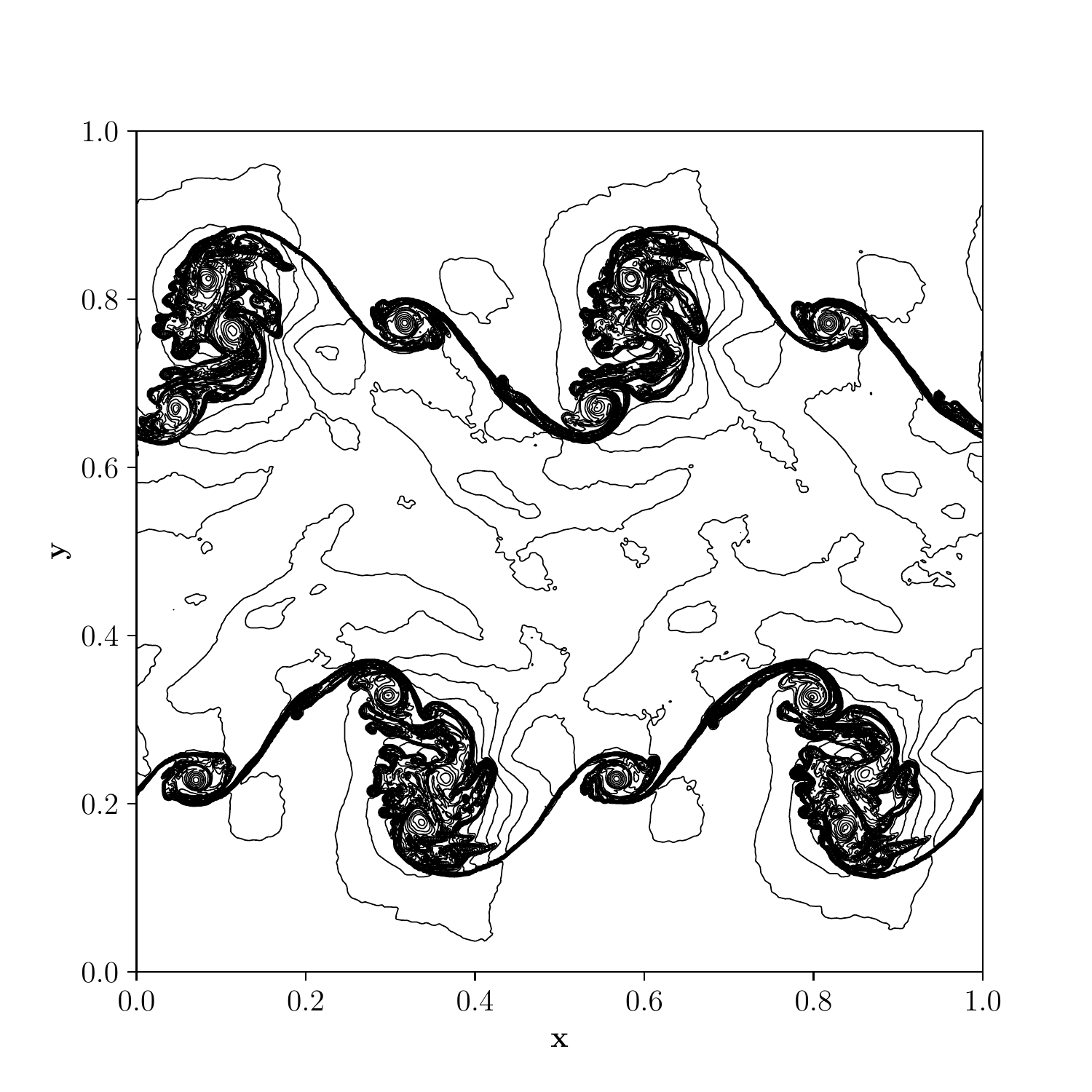}
        \label{fig:migb_khi}}
        \subfigure[MIG-S]{\includegraphics[width=0.3\textheight]{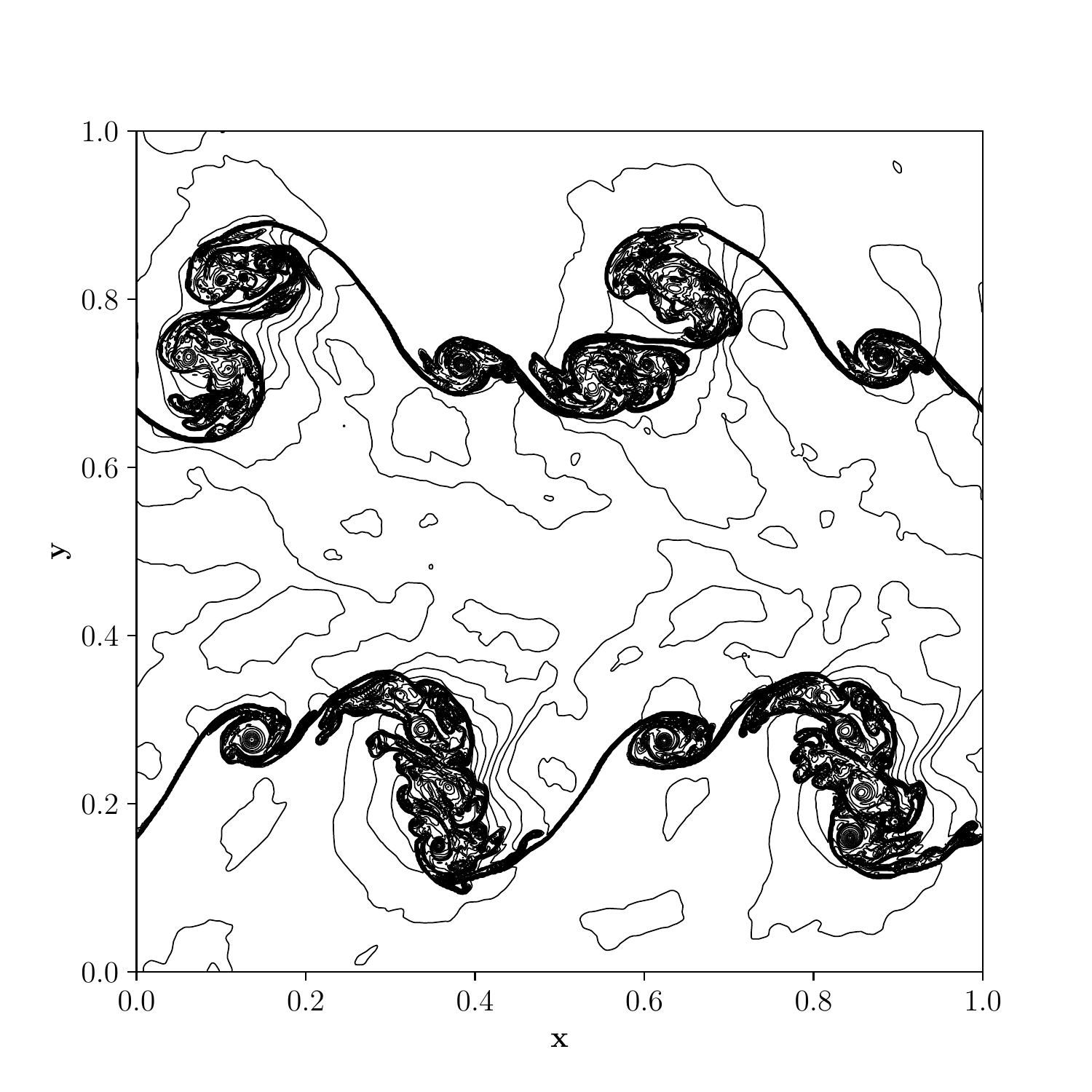}
        \label{fig:migs_khi}}
        \caption{Density contours of Example \ref{ex:kh} for the considered schemes. The figures are drawn with 30 contours. (a) TENO. (b) MIG-D. (c) MIG. (d) MIG-S.}
        \label{fig_KH}
\end{figure}

The periodic computational domain of $[x,y] = [0,1] \times [0,1]$ was discretized with $512 \times 512$ cells. The simulation was run until a final time of $t_f = 0.8$. The computed density iso-contours are shown in Fig. \ref{fig_KH}. All schemes captured complex small-scale vortices. However, the MIG-D scheme showed significant oscillations. In contrast, the MIG-S scheme was not only free of oscillations but also resolved the contact discontinuity significantly better (i.e., is thinner) than the MIG and TENO scheme.

\begin{example}\label{ex:SB}{Shock-Bubble interaction}
\end{example}

For the sixth example, we considered the shock-bubble (SB) interaction, \textcolor{black}{in which} a Mach 6 shock wave impacts a helium bubble \cite{Hu2010}. For simplicity, both air (the fluid of the shock wave) and helium were treated as ideal gases. The helium bubble was placed at $(x,y) = (0.25,0)$ within a domain of size $[x,y] = [0,1] \times [-0.5,0.5]$. The initial radius of the bubble was $0.15$. The shock front was initially placed at $x = 0.05$. A uniform grid size of $800 \times 800$ was used. Inflow and outflow conditions were applied at the left and right boundaries. Neumann boundary conditions with zero gradients for all variables were set at the top and bottom boundaries. The simulations were carried out with the shock-stable HLLC Riemann solver \cite{fleischmann2020shock} to prevent the carbuncle phenomenon. The initial conditions computed using exact Rankine-Hugoniot conditions were:

\begin{equation}
    (\rho, u, v, p)= 
    \begin{cases}
        (1, -3, 0, 1), & \text{pre-shocked air},
        \\
        (216/41, (1645/286) - 3, 0, 251/6),& \text{post-shocked air},
        \\
        (0.138, -3, 0, 1),& \text{helium bubble}.
    \end{cases}
    \label{sb}
\end{equation}

\begin{figure}[H]
        \centering\offinterlineskip
        \subfigure[TENO]{\includegraphics[width=0.48\textwidth]{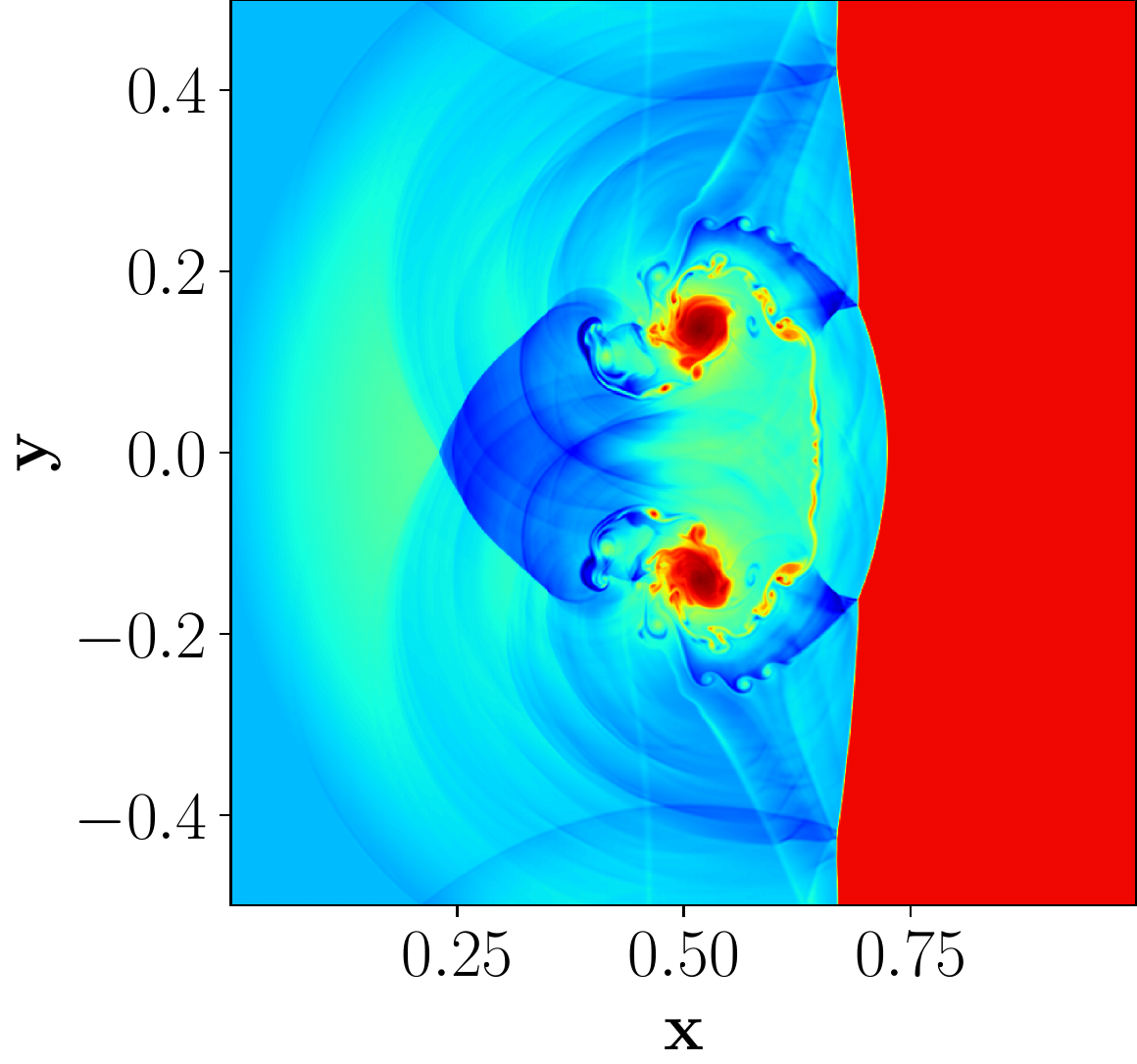}
        \label{fig:TDNO-Z_SB-LF}}
        \subfigure[MIG]{\includegraphics[width=0.48\textwidth]{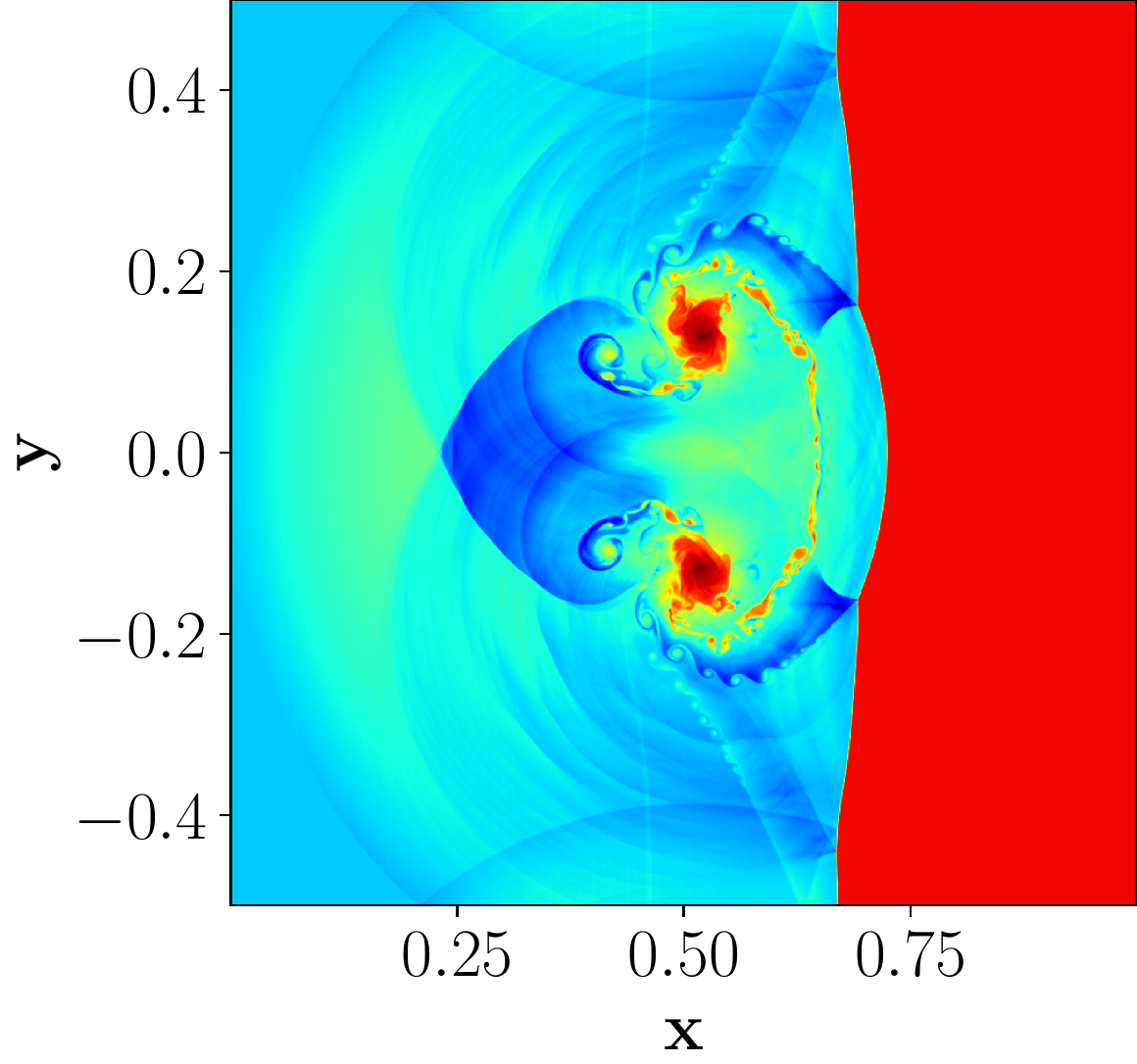}
        \label{fig:MIG_base-GLF}}
        \subfigure[MIG-S]{\includegraphics[width=0.48\textwidth]{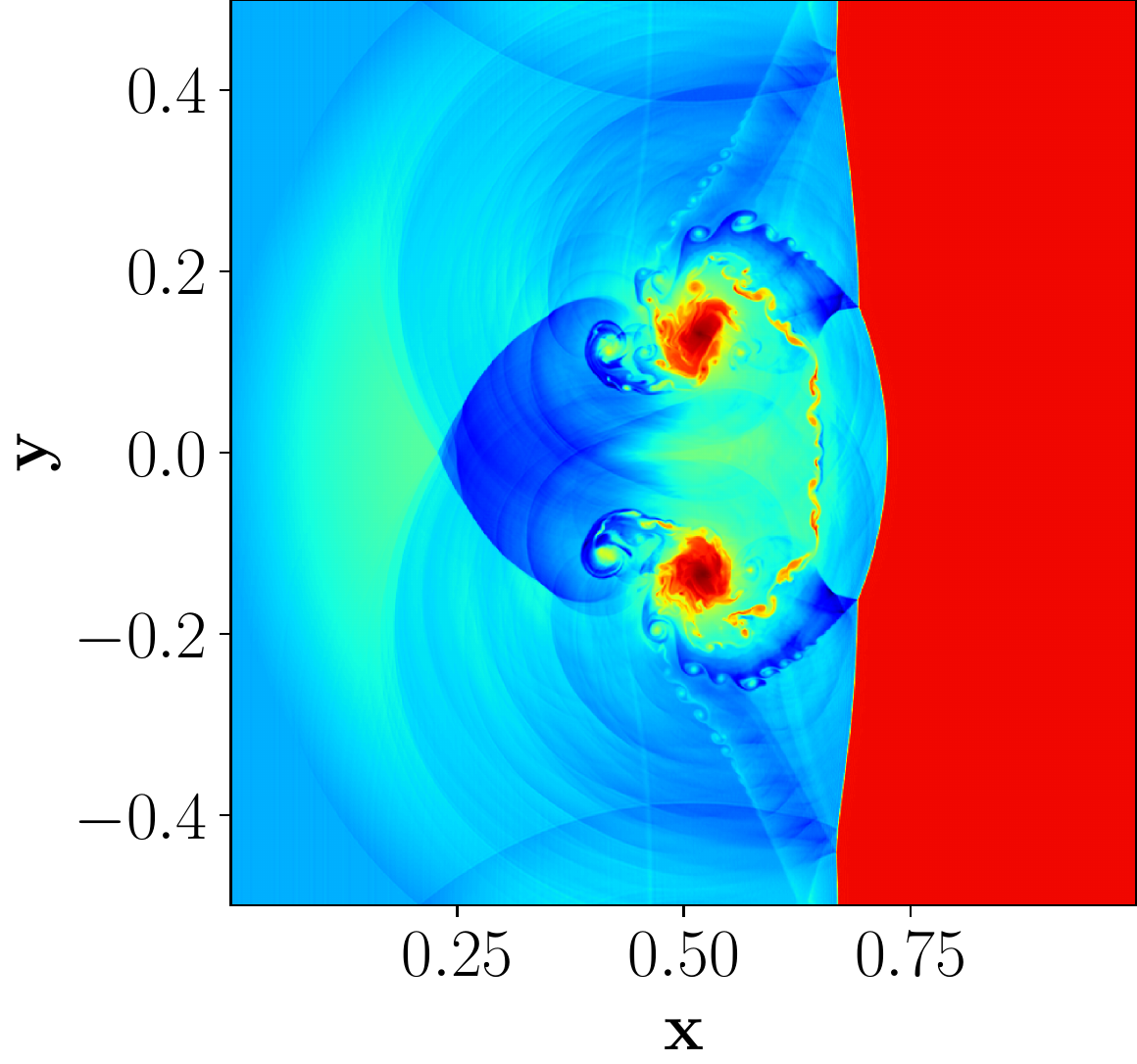}
        \label{fig:MIG_SB-LLFM}}
        \caption{Density contours of Example \ref{ex:SB} for the considered schemes. The contour of MIG-D is not included as it crashed. (a) TENO. (b) MIG. (c) MIG-S.}
        \label{fig_allbubble}
\end{figure}

Observing Fig. \ref{fig_allbubble}, the MIG-S scheme resolved more small-scale vortices than the TENO and MIG schemes. The MIG-D scheme failed within a few time steps for this test case and as a result, the contour was not included. The proposed method resolved the contact discontinuities and fine-scale structures and was also more computationally efficient than the MIG and TENO schemes as shown in Table \ref{tab:time}. \textcolor{black}{It is clear from Table \ref{tab:time} that MIG-D is computationally cheaper than the other schemes since the limiter is activated less often. However, it produced oscillatory results or crashed for some of the test cases as a result of it's inherent inability to detect contact discontinuities. On the other hand, MIG-S did not produce oscillatory results and was consistently computationally less expensive than MIG. Further, it was less expensive than the TENO scheme for some cases and produced superior resolution.}

\begin{table}[H]
    \centering
    \caption{Computational times of the considered schemes for the above examples.}
    \begin{tabular}{ c c c c c }
        \hline
        \hline
        Test case & MIG-D (s) & MIG (s) & MIG-S (s) & TENO (s) \\
        \hline
        Riemann & 6696                 & 8882  & 7207 & 6751 \\
        DMR     & 2222                 & 2939  & 2382 & 2103 \\
        RTI     & 1380 (oscillatory)   & 1448  & 1389 & 1510 \\
        RMI     & Crashed              & 373   & 351  & 316 \\
        KHI     & 1429 (oscillatory)   & 1740  & 1650 & 1620 \\
        SB      & Crashed              & 2247  & 1966 & 2210 \\
        \hline
        \hline
    \end{tabular}
    \label{tab:time}
\end{table}

It should also be noted that the TENO scheme does not have the advantage of sharing gradients between inviscid and viscous fluxes. Therefore, it is significantly more expensive than the MIG and MIG-S schemes for viscous flow simulations. Readers can refer to Chamarthi \cite{chamarthimig2022} for the detailed comparison of the MIG and TENO schemes with various viscous flux discretizations, which was shown to be important for a wide variety of flows \cite{chamarthi2022importance,chamarthi2023role}.

\begin{example}\label{ex:vs}{Viscous Shock tube}
\end{example}

Moving into viscous simulations, the Daru and Tenaud \cite{daru2009numerical} viscous shock-tube problem was used to demonstrate the advantage of MIG-S for the Navier-Stokes equations. The problem involves the propagation of a Mach 2.37 shock wave and contact discontinuity, which forms a thin boundary layer at the bottom wall. The shock wave interacts with this boundary layer, resulting in a complex vortex system, separation region, and a lambda-shaped shock pattern. This scenario makes it an ideal test case for evaluating high-resolution schemes. The case was run until a final time $t_f = 1$. The initial conditions are:

\begin{equation}
    (\rho, u, v, p)= 
    \begin{cases}
        (120, 0, 0, 120/\gamma), & \text{if } 0 < x < 0.5,
        \\
        (1.2, 0, 0, 1.2/\gamma),& \text{if } 0.5 \le x < 1.
    \end{cases}
    \label{vst}
\end{equation}

\begin{figure}[H]
    \centering
    \includegraphics[width=0.5\textwidth]{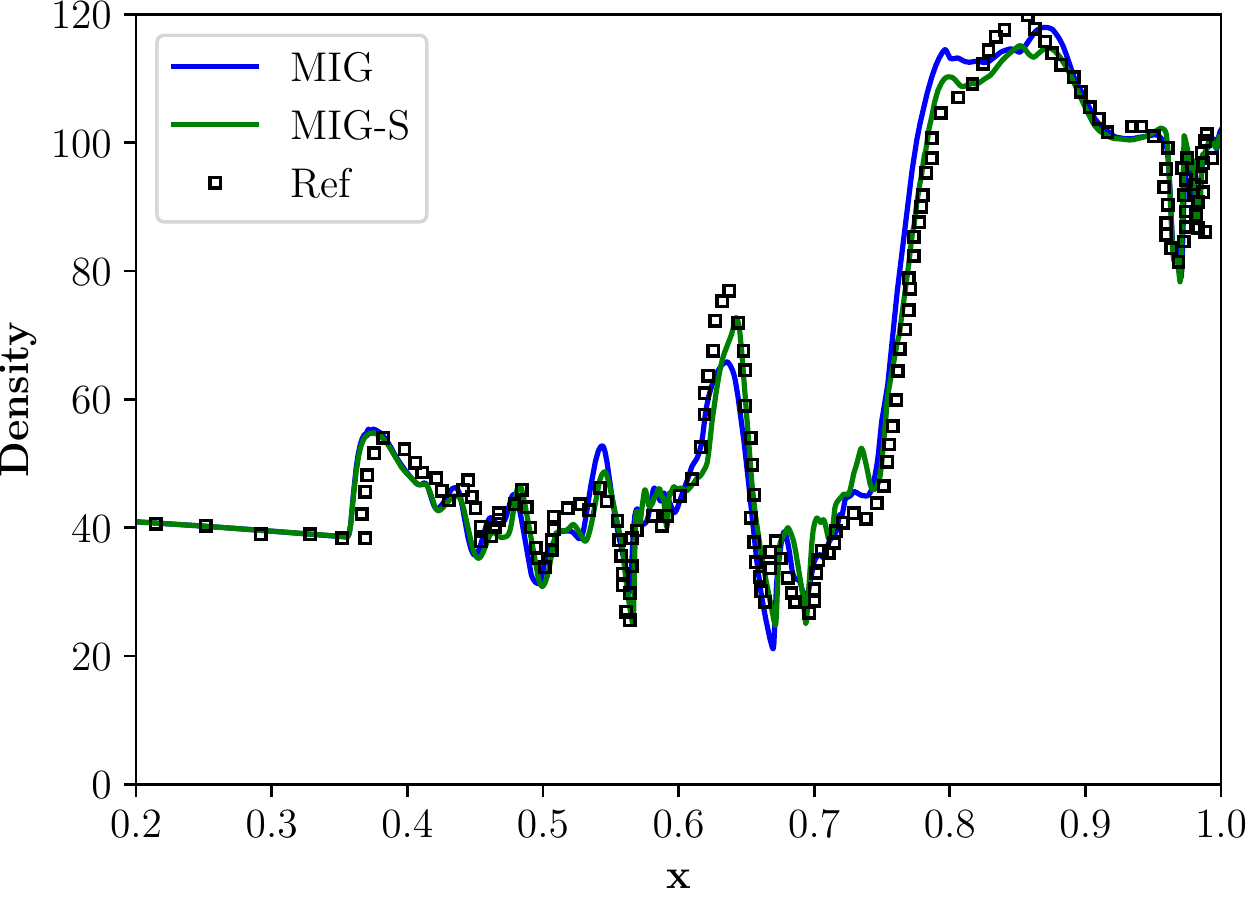}
    \caption{\textcolor{black}{Wall density comparison of MIG and MIG-S compared with the reference data of Kundu et al. \cite{kundu2021investigation}.}}
    \label{fig:vst_2500_line}
\end{figure}

The Reynolds number for this case was $\mathrm{Re} = 2500$. The computational domain was $[x,y] = [0,1] \times [0,0.5]$ and was uniformly discretized on a grid of $2000 \times 1000$. It is important to note that Kundu et al. \cite{kundu2021investigation} considered this case using a grid of 109 million cells. Even so, the present results obtained by both MIG and MIG-S are very close to their results despite using a grid 55 times smaller. Unfortunately, the TENO and MIG-D schemes failed for this test case, so the results are not presented (likewise in \cite{chamarthi2023implicit}). The superior performance of the proposed schemes compared to the TENO scheme highlights the importance of choosing appropriate numerical methods for simulating such flows. Additionally, the implicit gradient method has the advantage of sharing the gradients between inviscid and viscous fluxes (computed by Equation (\ref{eqn:compute_gradients})), which would not be possible with the TENO scheme. \textcolor{black}{Further, from Figure \ref{fig:vst_2500_line}, it is evident that MIG-S better matches the reference wall density data of Kundu et al. \cite{kundu2021investigation} compared with the MIG scheme. Additionally, MIG-S was approximately 25\% computationally cheaper than MIG for this case.}

\begin{figure}[H]
        \centering\offinterlineskip
        \subfigure[MIG, $t$ = 1]{\includegraphics[width=0.3\textheight]{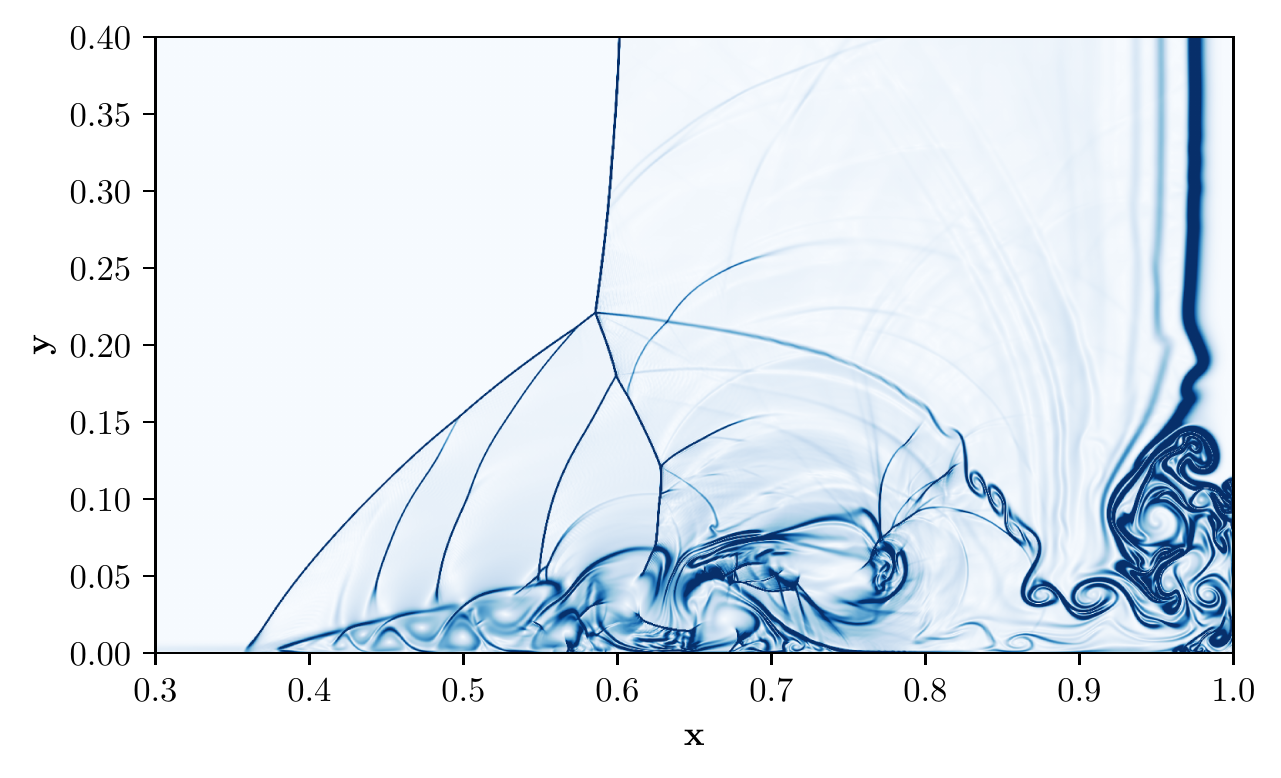}}
        \label{fig:IG6MP_VST_2500tp}
        \subfigure[MIG-S, $t$ = 1]{\includegraphics[width=0.3\textheight]{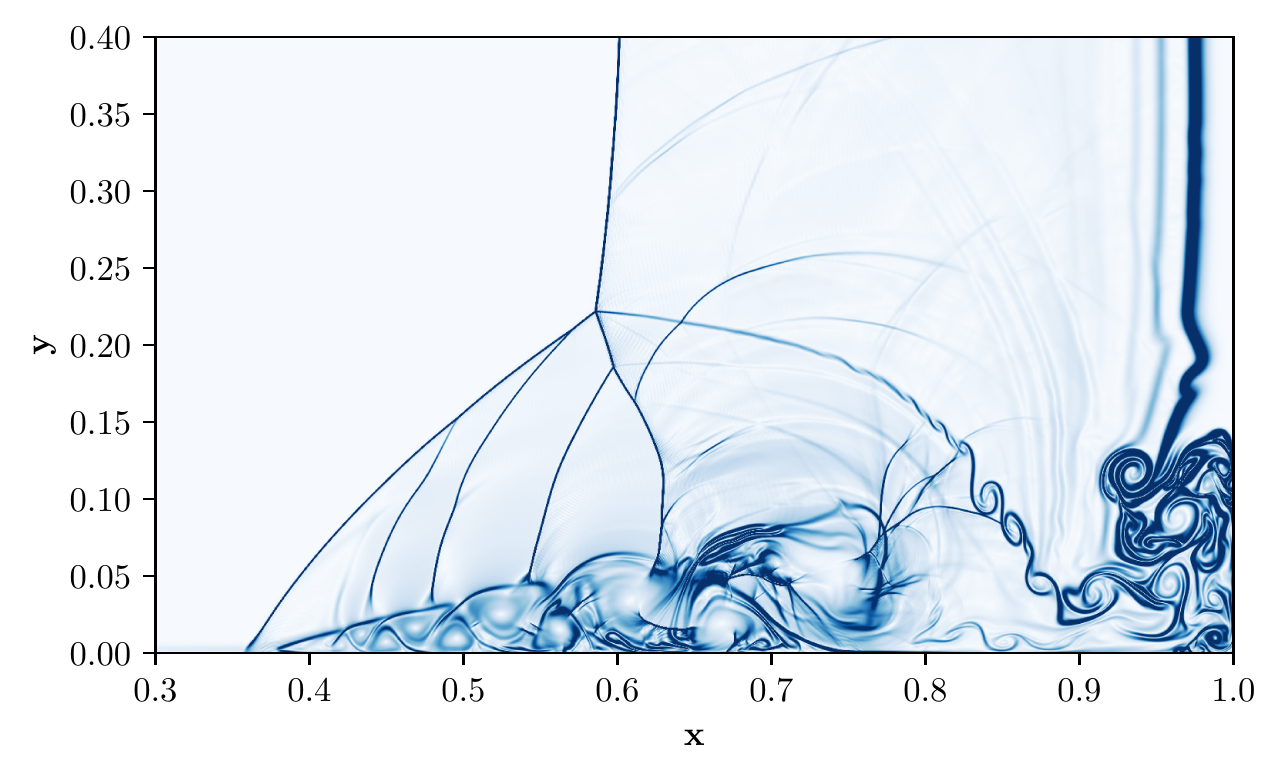}}
        \label{fig:IG4MP_VST_2500tp}
        \caption{Density gradient contours of Example \ref{ex:vs} for the considered schemes. The contours of MIG-D and TENO are not included as they crashed. (a) MIG. (b) MIG-S.}
        \label{fig_VST-2500}
\end{figure}

\begin{example}\label{ex:TGV}{Inviscid Taylor-Green Vortex}
\end{example}

For the final example, we investigated the performance of the considered schemes for solving the three-dimensional inviscid Taylor-Green vortex problem; a classical benchmark problem in computational fluid dynamics. \textcolor{black}{The case features the unsteady decay of a vortex. It is often used to assess the dissipation characteristics of numerical methods that solve the Euler equations: if kinetic energy loss is low and enstrophy generation is large, the method, in part, shows good dissipative properties.} The initial conditions for the simulation were:

\begin{equation}\label{itgv}
    \begin{Bmatrix}
        \rho \\
        u \\
        v \\
        w \\
        p \\
    \end{Bmatrix}
    =
    \begin{Bmatrix}
        1 \\
        \sin{x} \cos{y} \cos{z} \\
        -\cos{x} \sin{y} \cos{z} \\
        0 \\
        100 + \dfrac{\left[ \cos{(2z)} + 2 \right] \left[ \cos{(2x)} + \cos{(2y)} \right] - 2}{16}
    \end{Bmatrix}.
\end{equation}

The case was run in a periodic domain of size $x,y,z \in [0,2\pi)$ until $t_f = 10$ on a grid size of $64^3$, with a specific heat ratio of $\gamma = 5/3$. The case is considered incompressible since the mean pressure is significantly large. The study aimed to evaluate the ability of different schemes to preserve volume-averaged kinetic energy and enstrophy over time. Enstrophy, defined as the integral of the square of the vorticity, was used as a measure of the scheme's ability to preserve as many vortex structures as possible. The linear IG4H scheme \cite{chamarthi2023implicit,chamarthimig2022}, along with the nonlinear MIG, MIG-D, MIG-S, and TENO schemes, was also considered in this study. \textcolor{black}{The reason IG4-H was chosen was because it is a linear scheme with no discontinuity-detecting algorithm. As such, it should preserve the kinetic energy and enstrophy best.}

\begin{figure}[H]
        \centering\offinterlineskip
        \subfigure[Kinetic energy]{\includegraphics[width=0.45\textwidth]{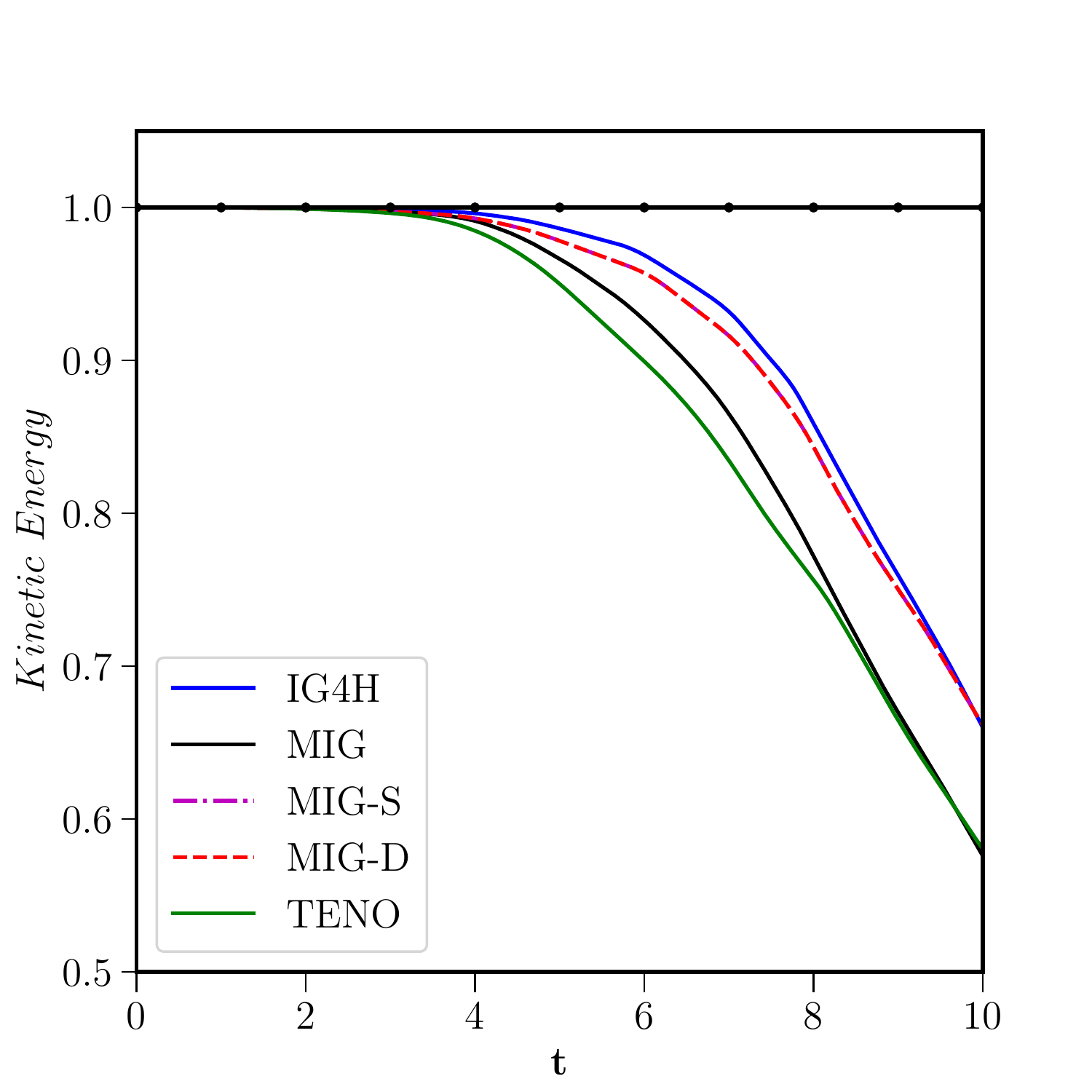}
        \label{fig:TGV_KE}}
        \subfigure[Enstrophy]{\includegraphics[width=0.45\textwidth]{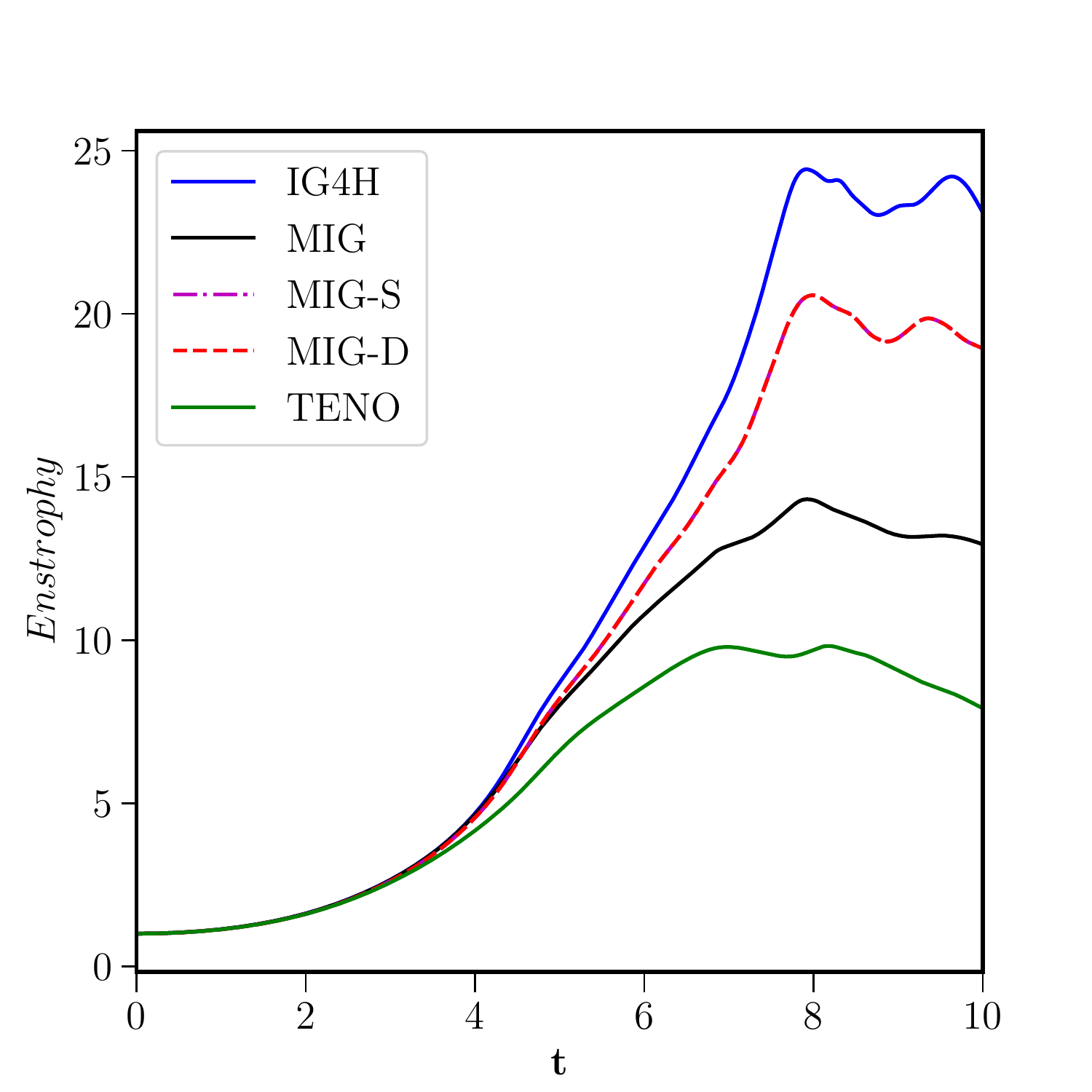}
        \label{fig:TGV_ens}}
        \caption{\textcolor{black}{Normalized}, volume-averaged kinetic energy and enstrophy of Example \ref{ex:TGV} for the considered schemes. Solid line with circles: exact solution; solid black line: MIG; solid green line: TENO; solid blue line: IG4H; dashed red line: MIG-D; dashed dotted magenta line: MIG-S. (a) Kinetic energy. (b) Enstrophy.}
        \label{fig_TGV}
\end{figure}

\begin{figure}[H]
        \centering\offinterlineskip
        \subfigure[TENO]{\includegraphics[width=0.48\textwidth]{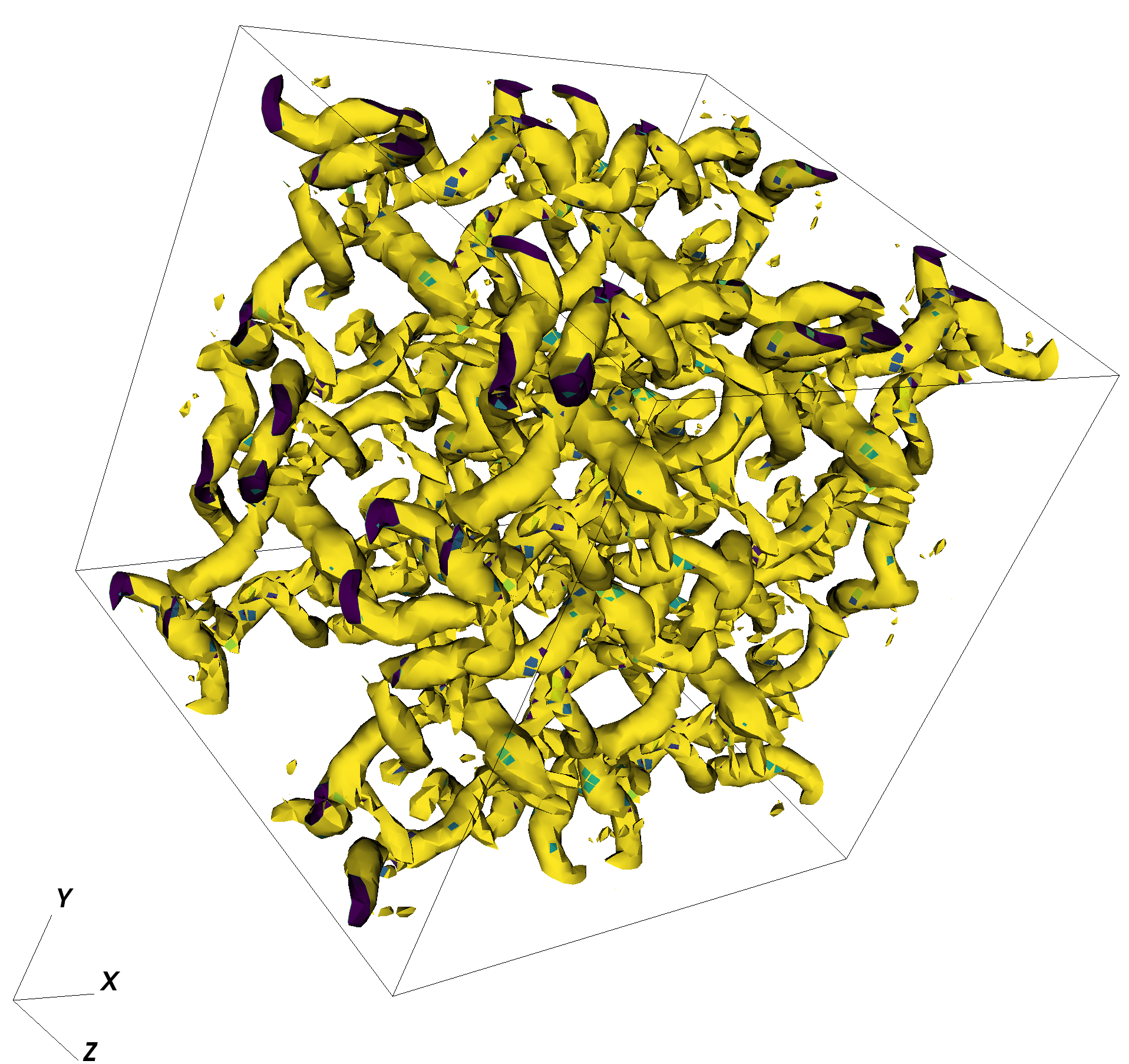}
        \label{fig:TDNO-Z_tgv-LF}}
        \subfigure[MIG]{\includegraphics[width=0.48\textwidth]{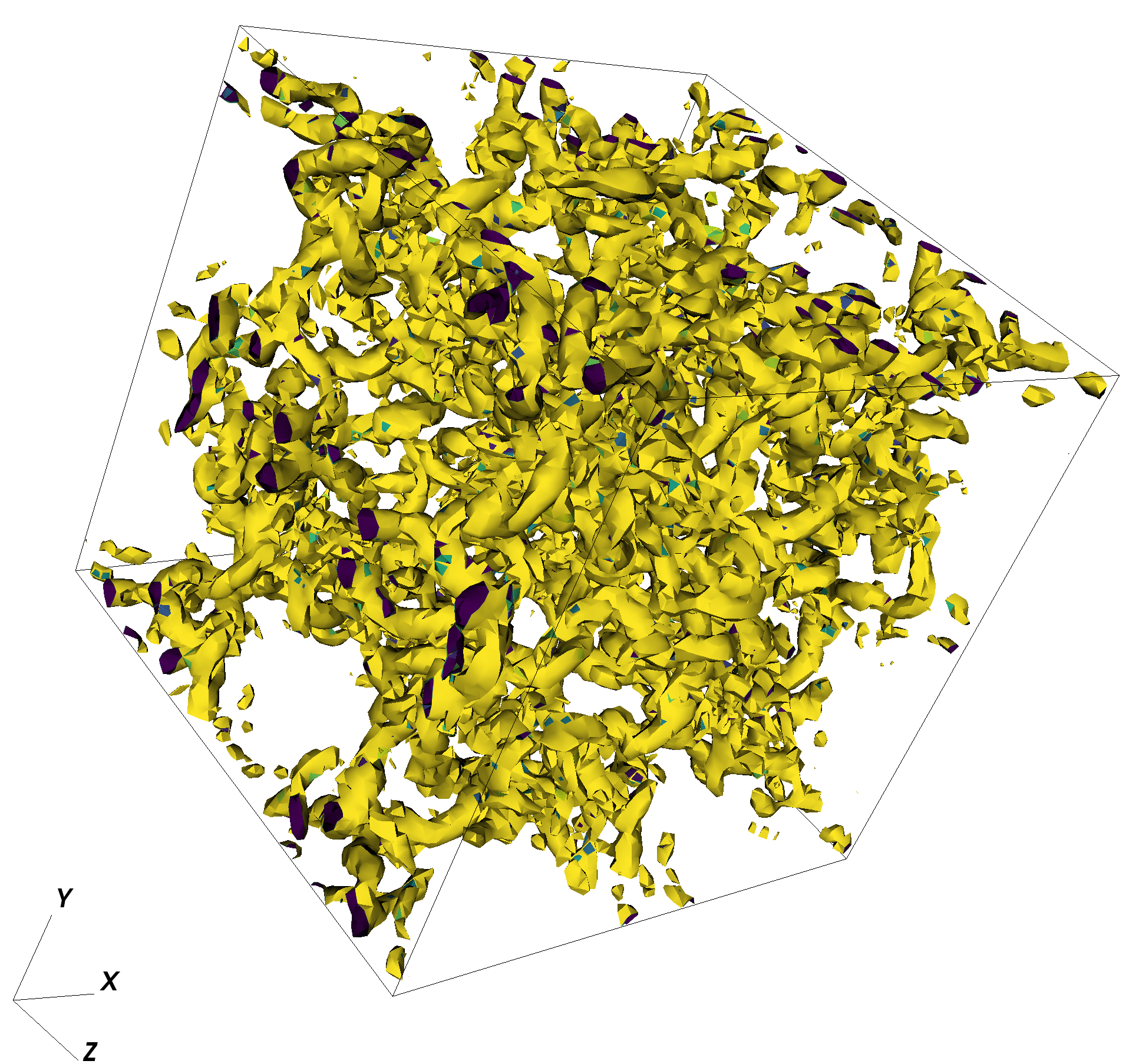}
        \label{fig:MIG_tgv-GLF}}
        \subfigure[MIG-S]{\includegraphics[width=0.48\textwidth]{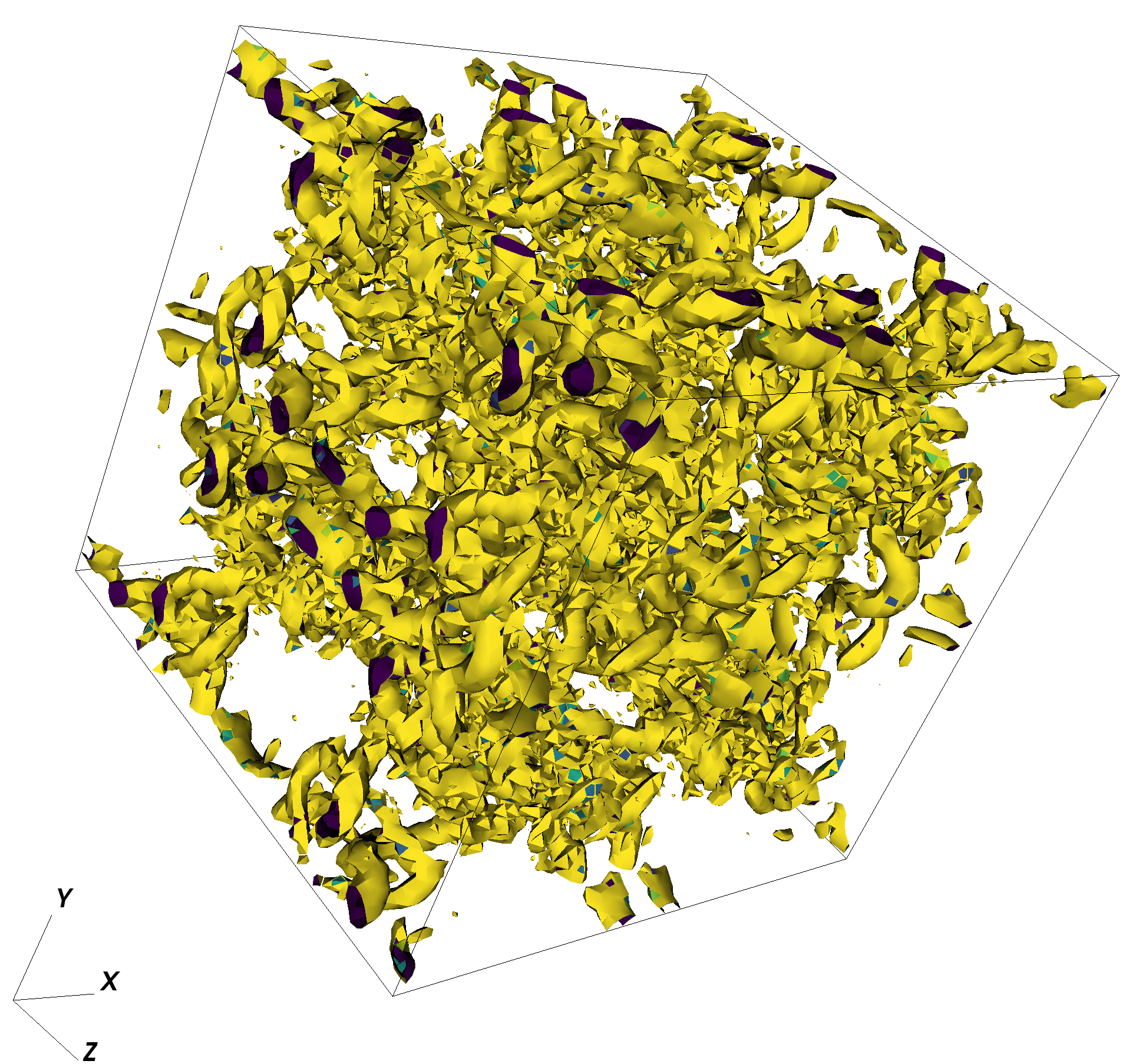}
        \label{fig:MIG-tgv-LLFM}}
        \caption{Resolved $Q = 2$ Q-criterion iso-surfaces of Example \ref{ex:TGV} for the considered schemes. (a) TENO. (b) MIG. (c) MIG-S.}
        \label{fig_qtgv}
\end{figure}

The kinetic energy evolution of all the numerical schemes is presented in Fig. \ref{fig:TGV_KE}. \textcolor{black}{As expected, IG4H preserves both kinetic energy and enstrophy better than the other considered schemes for the reasons cited above.} The results indicate that the MIG scheme better preserved kinetic energy than the TENO scheme. Furthermore, the MIG-D and MIG-S schemes preserved the kinetic energy significantly better than the MIG scheme due to the Ducros sensor. \textcolor{black}{This indicates that the MP limiting criterion is met far too often in the base MIG scheme compared with the MIG-D and MIG-S schemes, exhibiting the reason it may become too dissipative.} There is little-to-no difference between the MIG-S and MIG-D schemes, as this test case has no contact discontinuity. Furthermore, since the velocity gradients required to compute the enstrophy were already available, the proposed GBR method has an additional advantage. The MIG schemes (MIG, MIG-S, and MIG-D) performed the best in enstrophy preservation, as shown in Fig. \ref{fig:TGV_ens}. The resolved vortical structures from different schemes are shown in Fig. \ref{fig_qtgv}. The MIG-S scheme captured significantly more small-scale flow features than the MIG and TENO schemes.

 One may argue that very high-order TENO schemes \cite{Fu2016,fu2017targeted,fu2018new,fu2021very} may perform better, but it should also be noted that very-high-order TENO schemes are prone to oscillations and require a TVD or MP-like filter, as in \cite{li2021low}. Li et al. \cite{li2023family} mentioned that the smoothness indicators (which are a sum of \textbf{gradients} and are not re-used anywhere else) used in the ENO-type schemes for shock-capturing are very expensive and have proposed a regularization approach based on the MP criterion \cite{li2023family}.

\section{\label{sec:conclusions}Conclusions}

In this work, we proposed a novel selective discontinuity sensor approach that exploits the characteristic transformation that is commonly done for shock-capturing schemes. Since characteristic transformation already differentiates the characteristic waves in compressible flow systems, this can be used to appropriately treat various discontinuities. The novel sensor was applied to the MIG scheme of Chamarthi \cite{chamarthimig2022} and compared with the TENO scheme of Fu \cite{fu2019low}. We also considered the standard Ducros sensor applied to the MIG scheme. A suite of test cases \textcolor{black}{was} presented showing the efficacy of the proposed method in appropriately resolving shock waves and contact discontinuities while preserving important flow features as a result of it's low dissipation. Since the MIG family of schemes uses Gradient-Based Reconstruction, gradients can be re-used throughout the solver, resulting in increased computational efficiency. For test cases that did not include contact discontinuities, the results showed that the proposed sensor did not have an advantage over the standard Ducros sensor, which was expected. However, for cases with contact discontinuities, significantly better results were achieved. In some cases, the standard Ducros sensor produced oscillatory results and even crashed, whereas the proposed sensor showed no oscillations and successfully completed simulations with high resolution. The proposed sensor can also be used with other shock-capturing schemes, such as WENO, which already perform characteristic transformation. As such, it can readily be implemented into existing methods to significantly improve solution quality.

\begin{acknowledgments}

A.S. and N.H. were supported by the Technion Fellowship during this work.

\end{acknowledgments}

\appendix

\section*{\label{sec:appendixA}Appendix A}

\begin{figure}[H]
        \centering\offinterlineskip
        \subfigure[WENO-Z]{\includegraphics[width=0.2\textheight]{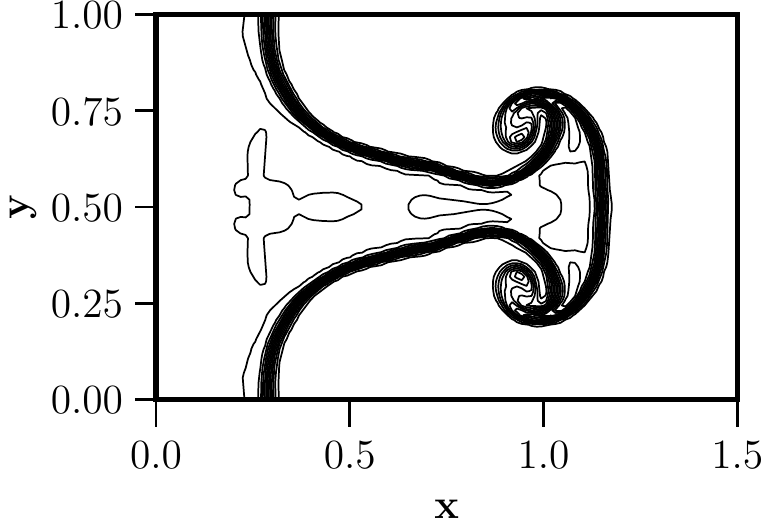}
        \label{fig:weno_rm}}
        \subfigure[WENO-Z-D]{\includegraphics[width=0.2\textheight]{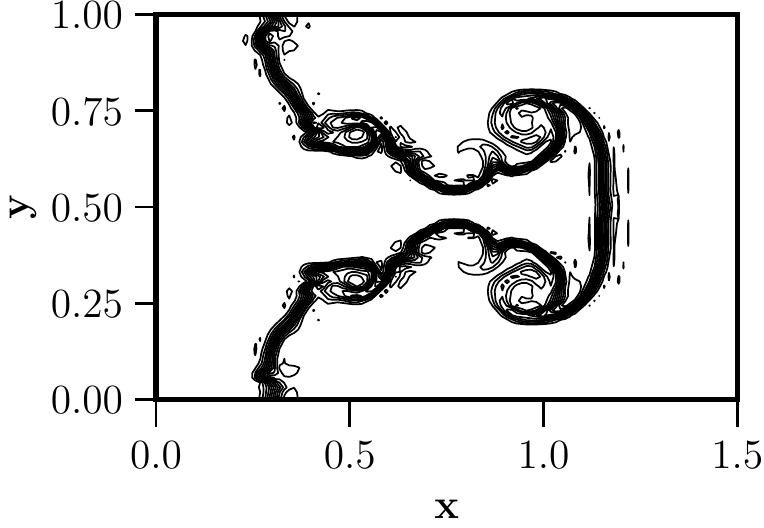}
        \label{fig:wenod_rm}}
        \subfigure[WENO-Z-S]{\includegraphics[width=0.2\textheight]{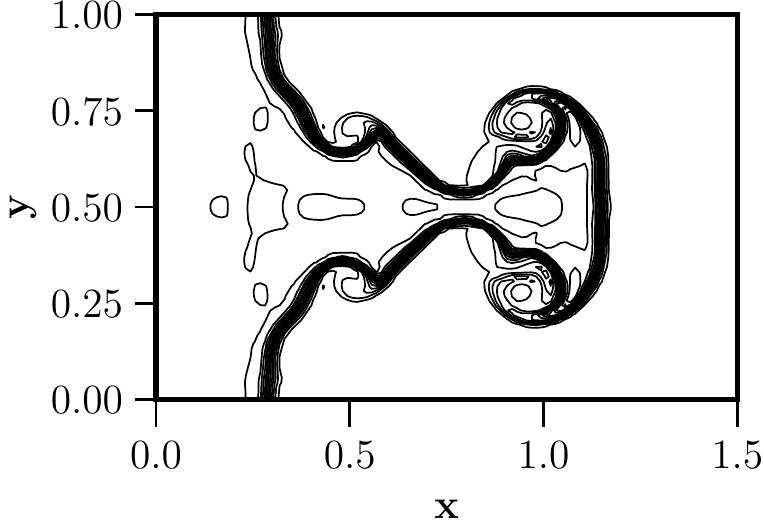}
        \label{fig:wenozs_rm}}
        \caption{Density contours of Example \ref{ex:rm} with the WENO-Z scheme, WENO-Z scheme with Ducros sensor, and the WENO-Z scheme with the selective sensor approach. (a) WENO-Z. (b) WENO-Z-D. (c) WENO-Z-S.}
        \label{fig_weno_rmi}
\end{figure}

\begin{figure}[H]
        \centering\offinterlineskip
        \subfigure[WENO-Z]{%
        \includegraphics[width=0.20\textwidth]{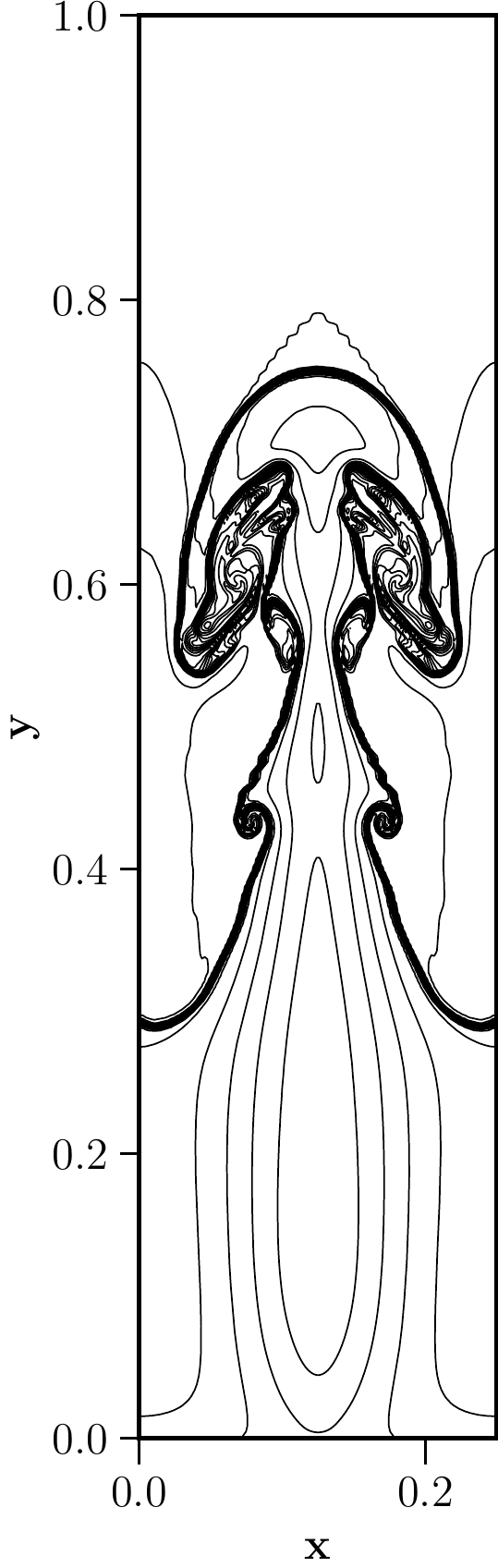}
        \label{fig:wen0_rt}}
        \subfigure[WENO-Z-D]{%
        \includegraphics[width=0.20\textwidth]{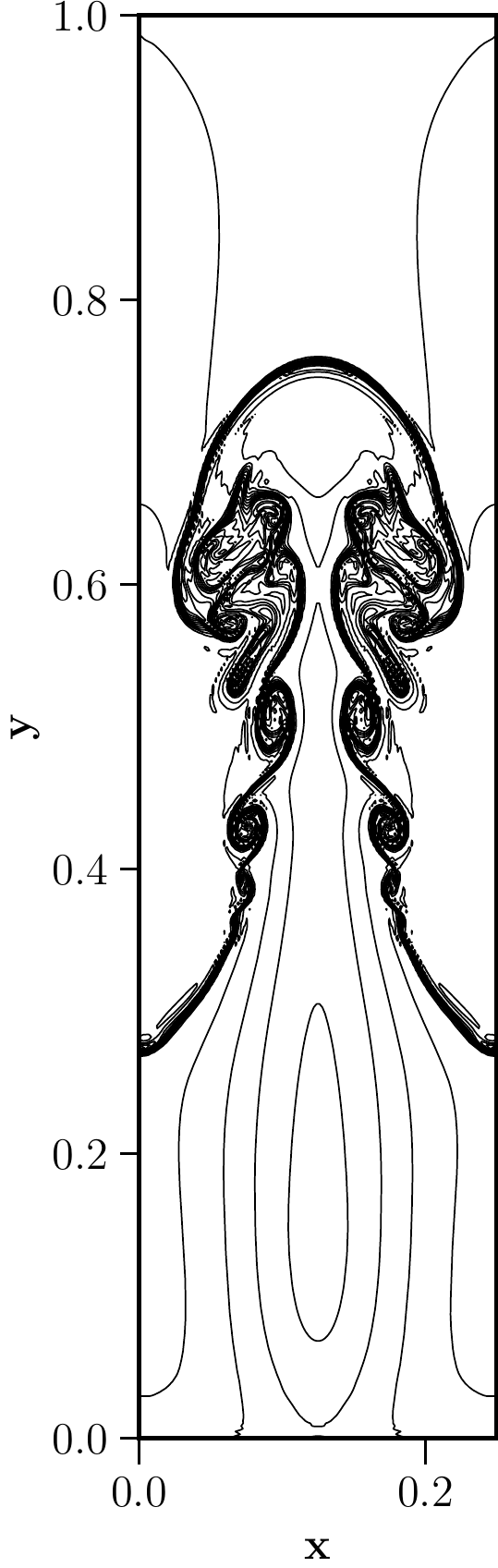}
        \label{fig:wenod_rt}}
        \subfigure[WENO-Z-S]{%
        \includegraphics[width=0.20\textwidth]{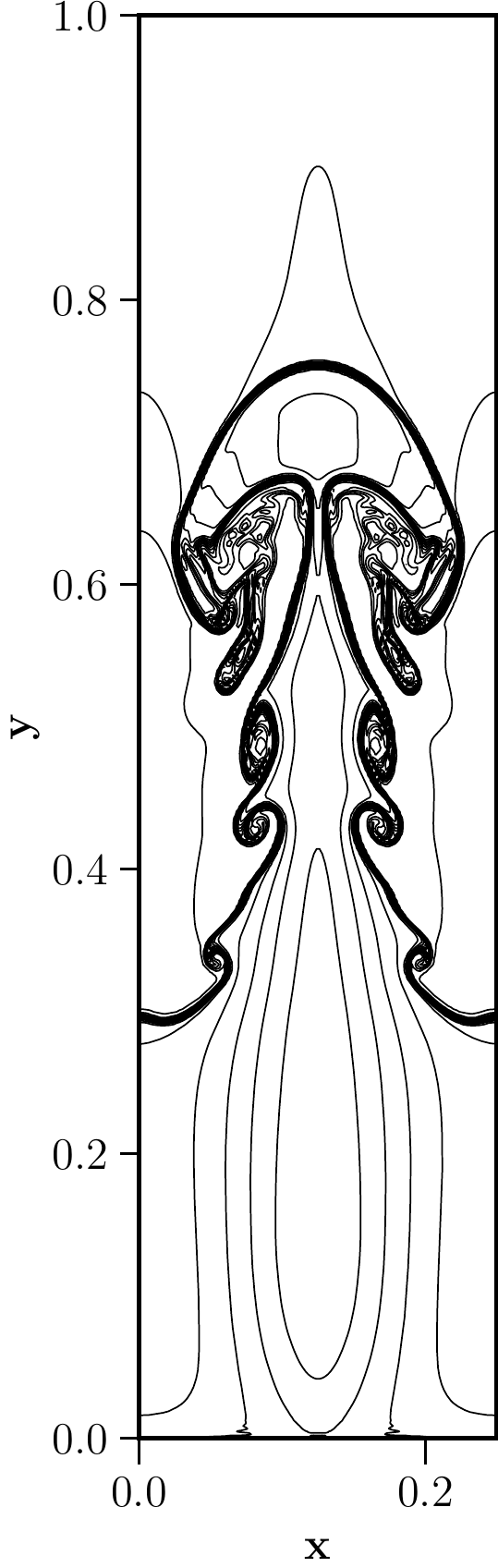}
        \label{fig:wenos_rt}}
        \caption{Density contours of Example \ref{ex:rm} with the WENO-Z scheme, WENO-Z scheme with Ducros sensor, and the WENO-Z scheme with the selective sensor approach. (a) WENO-Z. (b) WENO-Z-D. (c) WENO-Z-S.}
        \label{fig_weno_rt}
\end{figure}

We present the results obtained using the fifth-order WENO-Z scheme using the selective discontinuity detector approach. The WENO-Z scheme was implemented using conservative variables. The results for Examples \ref{ex:rm} and \ref{ex:rt} are shown in Figs. \ref{fig_weno_rmi} and \ref{fig_weno_rt}. It can be seen that the simulation carried out with the Ducros sensor alone, WENO-Z-D (Fig. \ref{fig:wenod_rm}), gave oscillatory results whereas the results obtained with the selective sensor approach, WENO-Z-S (Fig. \ref{fig:wenozs_rm}), is free of oscillations. The WENO-Z-S approach resolved the small-scale features better than the standard WENO-Z approach.

Similar observations can be made for Example \ref{ex:rt}. It can be seen that the simulation carried out with the Ducros sensor alone, WENO-Z-D (Fig. \ref{fig:wenod_rt}), gave oscillatory results whereas the results obtained with the selective sensor approach, WENO-Z-S (Fig. \ref{fig:wenos_rt}), is free of oscillations and better resolved the flow features in comparison with the standard WENO-Z scheme. These results indicate that the proposed selective discontinuity detector approach can also be used in conjunction with other methods, improving their solution quality significantly. Another important note is that the velocity gradients used in computing the Ducros sensor can also be re-used in the viscous flux discretization, thus making the approach efficient for viscous flow simulations.

\bibliographystyle{apsrev4-1}
\bibliography{aipsamp}

\end{document}